\documentclass[fleqn,usenatbib]{mnras}

\usepackage{newtxtext,newtxmath}

\usepackage[T1]{fontenc}
\usepackage{ae,aecompl}

\usepackage{graphicx}	
\usepackage{amsmath}	
\usepackage{amssymb}	

\usepackage{color}
\usepackage{ulem}

\newcommand{\VEC}{\mathbfit}

\newcommand{\hMpc}{\,h^{-1}\, {\rm Mpc}}
\newcommand{\hGpc}{\,h^{-1}\, {\rm Gpc}}

\newcommand{\hk}{\,h\,{\rm Mpc^{-1}}}

\definecolor{orange}{rgb}{1.0, 0.5, 0.0}
\definecolor{mygreen}{rgb}{0.0,0.5, 0.5}

\title[A decomposition formalism for the redshift-space bispectrum]
{A complete FFT-based decomposition formalism for the redshift-space bispectrum}

\author[N. S. Sugiyama et al.]{
Naonori S. Sugiyama,$^{1}$\thanks{E-mail: nao.s.sugiyama@gmail.com}
Shun Saito,$^{2}$ 
Florian Beutler,$^{3,4}$ 
and Hee-Jong Seo$^{5}$
\\
$^{1}$ Kavli Institute for the Physics and Mathematics of the Universe (WPI), \\
Todai Institutes for Advanced Study, The University of Tokyo, Chiba 277-8582, Japan\\
$^{2}$ Max-Planck-Institut f\"{u}r Astrophysik, Karl-Schwarzschild-Star{\ss}e 1, D-85740 Garching bei M\"{u}nchen, Germany\\
$^3$ Institute of Cosmology \& Gravitation, University of Portsmouth, Portsmouth, PO1 3FX, UK\\
$^4$ Lawrence Berkeley National Laboratory, 1 Cyclotron Road, Berkeley, CA 94720, USA\\
$^5$ Department of Physics and Astronomy, Ohio University, Clippinger Labs, Athens, OH 45701 
}

\date{}

\pubyear{}

\begin{document}
\label{firstpage}
\pagerange{\pageref{firstpage}--\pageref{lastpage}}
\maketitle

\begin{abstract}

To fully extract cosmological information from nonlinear galaxy distribution in redshift space, it is essential to include higher-order statistics beyond the two-point correlation function. In this paper, we propose a new decomposition formalism for computing the anisotropic bispectrum in redshift space and for measuring it from galaxy samples. Our formalism uses tri-polar spherical harmonic decomposition with zero total angular momentum to compress the 3D modes distribution in the redshift-space bispectrum. This approach preserves three fundamental properties of the Universe: statistical homogeneity, isotropy, and parity-symmetry, allowing us to efficiently separate the anisotropic signal induced by redshift-space distortions (RSDs) and the Alcock-Paczy\'{n}ski (AP) effect from the isotropic bispectrum. The relevant expansion coefficients in terms of the anisotropic signal are reduced to one multipole index $L$, and the $L> 0$ modes are induced only by the RSD or AP effects.
Our formalism has two advantages: (1) we can make use of Fast Fourier Transforms (FFTs) to measure the bispectrum; (2) it gives a simple expression to correct for the survey geometry, i.e., the survey window function. As a demonstration, we measure the decomposed bispectrum from the Baryon Oscillation Spectroscopic Survey (BOSS) Data Release 12, and, for the first time, present a $14\sigma$ detection of the anisotropic bispectrum in the $L=2$ mode.

\end{abstract}




\section{INTRODUCTION}
\label{Sec:Introduction}

In the current picture of structure formation, the standard theory of inflation~\citep{Starobinsky:1980te,Sato:1980yn,Guth:1980zm,Linde:1981mu,Albrecht:1982wi} predicts primordial density perturbations which are nearly Gaussian. If the density fluctuation of galaxies kept being a purely Gaussian random field during its evolution with cosmic time, the two-point function or its Fourier transform, the power spectrum, would completely characterize the spatial distribution of galaxies. However, the non-linear growth of structure produces strong non-Gaussian fluctuations through the mode-coupling of different scales, resulting in higher order statistics~\cite[e.g.,][]{Peebles1980lssu.book,Bernardeau:2001qr}. Consequently, cosmological information on the two-point statistics leaks to these non-Gaussian (non-linear) fluctuations. The three-point function or its Fourier space counterpart, the bispectrum, is thus a powerful tool for extracting a variety of important cosmological information on a given galaxy distribution, which complements that from the two-point statistics. 

With this motivation, there have been various efforts to measure the three-point statistics. The first measurements of the galaxy three-point function and the bispectrum were carried out in angular catalogues by~\cite{Peebles1975,Groth:1977gj} and~\cite{Fry:1982}, respectively. Subsequently, many analyses of the three-point statistics have been performed in spectroscopic galaxy surveys by~\cite{Kayo:2004fd,Jing:2004nb,Wang:2004km,Gaztanaga:2005an,Nichol:2006mg,Kulkarni:2007qu,Gaztanaga:2009sq,McBride:2011zn,McBride:2011zp,Marin:2011iv,Marin:2013bbb,Guo:2014lpa,Guo:2015iga,Slepian:2015hca} in configuration space, and by \cite{Scoccimarro:2001sp,Feldman:2001vk,Verde:2002sf,Gil-Marin:2015sta,Gil-Marin:2015baa} in Fourier space. Most of these studies were limited to certain choices of triangular configurations of the three-point statistics, because such measurements for all possible configurations have been computationally challenging. Over the past few years, with the progress of algorithm to measure the three-point function~\citep{Slepian:2015qza,Slepian:2016qwa} and the bispectrum~\citep{Scoccimarro:2015bla}, one has started to make use of the information of full triangular configurations on the three-point statistics to constrain cosmological parameters in configuration space~\citep{Slepian:2017kfz} and Fourier space~\citep{Gil-Marin:2017wya,Pearson:2017wtw}. 

There remains one important issue in measuring the three-point statistics in the galaxy redshift surveys: all of the previous analyses measure \textit{only one component}, i.e., the monopole, after spherically averaging over the line-of-sight (LOS) direction. The observed clustering pattern of galaxies is anisotropically distorted by the peculiar velocities of galaxies along the LOS, known as redshift-space distortions (RSDs; see~\citealt{Hamilton:1997zq} for a review). An additional anisotropic signal arises due to the conversion from the observed redshifts into radial distances with incorrect cosmological parameters, which is known as the Alcock-Paczy\'{n}ski (AP) effect~\citep{Alcock:1979mp}. 
These two effects, the RSD, and AP effects, each leave their distinctive mark on the observed galaxy clustering.

In the case of the power spectrum which depends only on one wave vector, $\VEC{k}$, the anisotropic distortion is quantified by an angle between the wave vector $\VEC{k}$ and the LOS unit vector $\hat{n}$ and thus can be entirely decomposed using the Legendre polynomials ${\cal L}_{\ell}(\hat{k}\cdot\hat{n})$, where ${\cal L}_{\ell}$ denotes the Legendre polynomial at $\ell$-th order. Note that the LOS vector is locally defined with respect to each galaxy pair (local plain-parallel approximation \cite{Beutler:2013yhm}). In the bispectrum case, because of its three angular dependences, $\hat{k}_1$, $\hat{k}_2$, and $\hat{n}$, one can arbitrarily choose a coordinate system to characterize its LOS dependence. \cite{Scoccimarro:1999ed} selected $\hat{k}_1$ as the $z$-axis and decomposed the LOS dependence into spherical harmonics: $B(\VEC{k}_1,\VEC{k}_2,\hat{n}) = \sum_{LM}B_{LM}(\VEC{k}_1,\VEC{k}_2) Y_L^M(\hat{n})$. \cite{Hashimoto:2017klo} defined $(\hat{k}_1\times \hat{k}_2)$ as the $z$-axis in the same bispectrum decomposition.
More recently, \cite{Slepian:2017lpm} proposed in configuration space to take the LOS as the $z$-axis and to expand the three-point function into the product of two spherical harmonics: $\xi(\VEC{r}_1,\VEC{r}_2,\hat{n}) = \sum_{\ell_1\ell_2 m}\xi_{\ell_1\ell_2}^{m}(r_1,r_2,\hat{n}) Y_{\ell_1}^{m}(\hat{r}_1) Y_{\ell_2}^{m*}(\hat{r}_2)$, where these spherical harmonic functions have the same $m$ mode (spin). We stress here that all of these expansion coefficients include the $M\neq0$ ($m\neq0$) modes and hence are variable under rotations. As the Universe is thought to be statistically homogeneous, isotropic, and parity symmetric, any decomposition which preserves these fundamental properties, can significantly simplify the analysis.

The main goal of this paper is therefore to propose a more efficient way to distinguish the anisotropic signal on the bispectrum (and the three-point function) induced by the RSD or AP effects from the isotropic signal. To reach the goal, we present a new decomposition of the bispectrum into the a tri-polar spherical harmonic (TripoSH; \citealt{Varshalovich1988qtam.book}) basis $\{\{Y_{\ell_1}(\hat{k}_1)\otimes Y_{\ell_2}(\hat{k}_2)\}_{\ell_{12}}\otimes Y_{L}(\hat{n})\}_{JM_{J}}$, which is a tensor product of spherical harmonics with three different arguments. The three-point function can also be decomposed into a similar basis: $\{\{Y_{\ell_1}(\hat{r}_1)\otimes Y_{\ell_2}(\hat{r}_2)\}_{\ell_{12}}\otimes Y_{L}(\hat{n})\}_{JM_{J}}$. In this basis, the assumption of statistical isotropy allows only the $J=0$ mode, i.e., zero total angular momentum. Furthermore, the parity symmetry assumption restricts allowed multipoles to $\ell_1+\ell_2+L={\rm even}$. This kind of basis was previously utilized to deal with the angular bispectrum of the cosmic microwave background (CMB) radiation (for the latest results, see~\citealt{Ade:2013ydc,Ade:2015ava} and references therein) and the wide-angle effect in large-scale structure~\cite[e.g.,][]{Szapudi:2004gh}. The bispectrum is then characterized by two wavenumbers, $k_1$ and $k_2$, and three multipoles, $\ell_1$, $\ell_2$, and $L$, where we use an upper-case index $L$ for anisotropic distortions along the LOS. We can then single out only the anisotropic signal by computing the $L\neq 0$ modes.

The formalism has two advantages. First, we can make use of Fast Fourier Transform (FFT) to measure our bispectrum multipoles by extending the \citet{Scoccimarro:2015bla} estimator to ours. For the three-point function multipoles, our estimator is closely related to that presented by~\citet{Slepian:2016qwa}. Second, we can correct for survey geometry effects on the bispectrum/three-point function in a similar manner to the \citet{Wilson:2015lup, Beutler:2016arn} formalism, which has been developed for the power spectrum analysis. 

As a demonstration of the efficiency of our estimators, we measure the bispectrum multipoles from a publicly available galaxy sample, the Baryon Oscillation Spectroscopic Survey Data Release 12 (BOSS DR12;~\citet{Alam:2015mbd})~\footnote{\url{http://www.sdss.org/dr12/}} and investigate the statistical properties of the measured bispectrum multipoles by computing their covariance matrices and signal-to-noise ratios. The covariance matrix shows an interesting feature that the diagonal elements (i.e., $k_1=k_2$) of the bispectrum multipoles estimated at different scales, i.e., at $k_1\, (=k_2)$ and $k'_1\, (=k'_2)$ for $k_1\neq k_1'$, are nearly uncorrelated like in the case of the power spectrum. We further find that for each of the $L=0$ and $2$ modes, the lowest order of the bispectrum multipoles yields the highest signal-to-noise ratio in the $L$ mode. These findings can significantly simplify the bispectrum analysis for such galaxy samples. While we leave a detailed analysis of the redshift-space bispectrum for future work, we will, for the first time, present a $14\sigma$ detection of the lowest order $L=2$ mode and a $6\sigma$ detection of the next leading order $L=2$ mode, which are induced only by anisotropic signals on the bispectrum (i.e. the RSD or AP effects).

The outline of the paper is as follows. In Section~\ref{Sec:NewDecompositionFormalsim} we present the new decomposition formalism of the bispectrum and three-point function to characterize the RSD effect on them. In Section~\ref{Sec:ComparisonWithPreviousWorks} we provide a comparison with previous works. In Section~\ref{Sec:Estimator} we describe the estimators to measure our bispectrum and three-point function multipoles. In Section~\ref{Sec:Measurements} we demonstrate our formalism by measuring the bispectrum multipoles of the BOSS DR12 galaxy sample and compute their covariance matrices and signal-to-noise ratios. In Section~\ref{Sec:SurveyWindowFunctions} we show how to correct for the survey geometry effect when theoretically modeling the observed bispectrum and three-point function multipoles. We present a summary and conclusions in Section~\ref{Sec:Conclusion}. We provide six  appendixes for further clearification and systematics test: in Appendix~\ref{Ap:UsefulIdentities} we summarize the identities used for the derivations of the equations used in this paper; in Appendix~\ref{Ap:LOS} we investigate the dependence of how to choose the LOS direction on the bispectrum measurements; in Appendix~\ref{Ap:MassAssignmentFunction} we provide a comparison of several interpolation schemes of density fields; in Appendix~\ref{Ap:FKPWeight} we show how the FKP weighting works in our decomposition formalism; in Appendix~\ref{Ap:HartlapAndPercival} we describe the Hartlap factor and the Percival factor; and in Appendix~\ref{Ap:WindowCorrections} we detail the bispectrum multipoles including survey window corrections. 

\section{NEW DECOMPOSITION FORMALISM}
\label{Sec:NewDecompositionFormalsim}

The three-point function and its Fourier transform, the bispectrum, are potential tools for extracting cosmological information that leaks to non-Gaussian fluctuations of galaxy clustering. To put these in context, we compute the number density of galaxies, $n_{\rm g}$, in fractional units relative to the background density, $\bar{n}_{\rm g}$: $n_{\rm g} = \bar{n}_{\rm g}\left( 1 + \delta \right)$, where $\delta$ is the density contrast. The three-point function is then defined as $\left\langle\delta(\VEC{x}_1)\delta(\VEC{x}_2)\delta(\VEC{x}_3)\right\rangle$, which is the ensemble average of the product of the three density contrasts at points $\VEC{x}_1$, $\VEC{x}_2$, and $\VEC{x}_3$.

The Universe is thought to be statistically homogeneous, and hence, the three-point function can be characterized by two relative coordinates, $\VEC{r}_1 = \VEC{x}_1-\VEC{x}_3$ and $\VEC{r}_2 = \VEC{x}_2-\VEC{x}_3$: 
\begin{eqnarray}
	 \zeta(\VEC{r}_1,\VEC{r}_2) = \left\langle\, \delta(\VEC{x}_1)\delta(\VEC{x}_2)\delta(\VEC{x}_3)\,\right\rangle.
\end{eqnarray}
Using the Fourier transform of the density contrast, $\delta(\VEC{k}) = \int d^3x\, {\rm e}^{-{\rm i}\VEC{k}\cdot\VEC{x}}\,\delta(\VEC{x})$, we define the bispectrum as
\begin{eqnarray}
	\left\langle \delta(\VEC{k}_1)\delta(\VEC{k}_2)\delta(\VEC{k}_3) \right\rangle
	=(2\pi)^3\delta_{\rm D}\left( \VEC{k}_1+\VEC{k}_2+\VEC{k}_3 \right) B(\VEC{k}_1,\VEC{k}_2),
\end{eqnarray}
where $\delta_{\rm D}$ represents the Dirac delta function. In the bispectrum case, its statistical homogeneity corresponds to the triangle condition, $\VEC{k}_1+\VEC{k}_2+\VEC{k}_3=0$.

The observed position of galaxies $\VEC{x}$ is displaced from the real-space galaxy position $\VEC{x}_{\rm r}$ by the physical peculiar velocity of galaxies $\VEC{v}$ along the LOS direction:
\begin{eqnarray}
	\VEC{x} = \VEC{x}_{\rm r} + \frac{\VEC{v}(\VEC{x}_{\rm r})\cdot\hat{x}}{aH(a)} \hat{x},
\end{eqnarray}
where $a$ is the scale factor, $H(a)$ is the Hubble expansion parameter and $\hat{x}=\VEC{x}/|\VEC{x}|$ is a unit vector pointing to the galaxy from the origin. The observed galaxy density is then distorted along the LOS direction, the so-called RSD effect. It is commonly assumed that the anisotropic distortion of the galaxy clustering is characterized by only one global LOS direction $\hat{n}$, the so-called \textit{global} plane-parallel approximation: $\zeta(\VEC{r}_1,\VEC{r}_2,\hat{n})$ and $B(\VEC{k}_1,\VEC{k}_2,\hat{n})$. However, this assumption does not hold for actual galaxy data, because observed galaxy positions have their own LOS directions, and the three-point function should depend on three LOS directions, $\hat{x}_1$, $\hat{x}_2$, and $\hat{x}_3$, in the most general case. Throughout this paper, we apply the \textit{local} plane-parallel approximation, $\hat{x}_1\approx\hat{x}_2\approx\hat{x}_3$, and choose $\hat{x}_3$ as the LOS direction for the triangular configuration formed by $\VEC{x}_1$, $\VEC{x}_2$, and $\VEC{x}_3$ when measuring the bispectrum from a galaxy sample in Section~\ref{Sec:Estimator}. To validate this choice, we show in Appendix~\ref{Ap:LOS} that the difference among the bispectrum measurements choosing each of $\hat{x}_1$, $\hat{x}_2$, and $\hat{x}_3$ as the LOS is negligibly small in the BOSS analysis. We will discuss the modeling of the bispectrum under the \textit{local} plane-parallel approximation by taking into account survey geometry effects in Section~\ref{Sec:SurveyWindowFunctions}.

The three angular-dependences in the bispectrum, $\hat{k}_1$, $\hat{k}_2$, and $\hat{n}$, can be generally decomposed into spherical harmonics $Y_{\ell}^m$:
\begin{eqnarray}
	B(\VEC{k}_1,\VEC{k}_2,\hat{n}) &=&  \sum_{\ell_1\ell_2 L} \sum_{m_1m_2M}B_{\ell_1\ell_2L}^{m_1m_2M}(k_1,k_2) \nonumber \\
	&\times&  y_{\ell_1}^{m_1}(\hat{k}_1) y_{\ell_2}^{m_2}(\hat{k}_2) y_{L}^{M}(\hat{n}),
\end{eqnarray}
where $y_{\ell}^m = \sqrt{4\pi/(2\ell+1)}\, Y_{\ell}^m$ is a normalized spherical harmonic function, and the corresponding expansion coefficients are given by
\begin{eqnarray}
	B_{\ell_1\ell_2L}^{m_1m_2M}(k_1,k_2) 
	\hspace{-0.23cm}&=&\hspace{-0.23cm}
	N_{\ell_1\ell_2L} \int \frac{d^2\hat{k}_1}{4\pi}\int \frac{d^2\hat{k}_2}{4\pi}\int \frac{d^2\hat{n}}{4\pi} \nonumber \\
	&\times& y^{m_1*}_{\ell_1}(\hat{k}_1) y^{m_2*}_{\ell_2}(\hat{k}_2)y^{M*}_{L}(\hat{n}) B(\VEC{k}_1,\VEC{k}_2,\hat{n}),
	\nonumber \\
	\label{Eq:SH_coefficients}
\end{eqnarray}
with $N_{\ell_1\ell_2L} = (2\ell_1+1)(2\ell_2+1)(2L+1)$. Throughout this paper, we use upper-case indices $L,M$ for the expansion with respect to the angles relevant to LOS. Namely, the modes with $L>0$ are induced by the RSD or AP effects.

Alternatively we can decompose the bispectrum using TripoSH expansion~\citep{Varshalovich1988qtam.book}:
\begin{eqnarray}
	B(\VEC{k}_1,\VEC{k}_2,\hat{n})
	\hspace{-0.25cm}&=&\hspace{-0.25cm}
	\sum_{JM_{J}}\sum_{\ell_1\ell_2L \ell_{12}} B_{\ell_1\ell_2 \ell_{12}L}^{J M_{J}}(k_1,k_2) S_{\ell_1\ell_2\ell_{12}L}^{J M_J}(\hat{k}_1,\hat{k}_2,\hat{n}),
    \nonumber \\
	\label{Eq:TripoSH}
\end{eqnarray}
where we defined a normalized TripoSH basis as
\begin{eqnarray}
	S_{\ell_1\ell_2\ell_{12}L}^{J M_J}(\hat{k}_1,\hat{k}_2,\hat{n})
   \hspace{-0.25cm} &=& \hspace{-0.25cm}
    \sum_{m_1m_2m_{12}M} C_{\ell_1 m_1;\ell_2m_2}^{\ell_{12}m_{12}} C_{\ell_{12} m_{12};LM}^{J M_J} \nonumber \\
   \hspace{-0.25cm} &\times& 
   y_{\ell_1}^{m_1}(\hat{k}_1) y_{\ell_2}^{m_2}(\hat{k}_2) y_L^M(\hat{n}),
\end{eqnarray}
and the TripoSH coefficients are then given by
\begin{eqnarray}
	 B_{\ell_1\ell_2\ell_{12}L}^{JM_J}(k_1,k_2)
    \hspace{-0.25cm} &=& \hspace{-0.25cm}
    \sum_{m_1m_2m_{12}M} C_{\ell_1 m_1;\ell_2m_2}^{\ell_{12}m_{12}} C_{\ell_{12} m_{12};LM}^{J M_J} \nonumber \\
   \hspace{-0.25cm} &\times&   B_{\ell_1\ell_2L}^{m_1m_2M}(k_1,k_2),
\end{eqnarray}
with the Clebsch-Gordan coefficients 
\begin{eqnarray}
C_{\ell_1m_1;\ell_2 m_2}^{\ell_3 m_3}=(-1)^{\ell_1-\ell_2+m_3} \sqrt{2\ell_3+1}  \left( \begin{smallmatrix} \ell_1 & \ell_2 & \ell_3\\ m_1 & m_2 & -m_3\end{smallmatrix}  \right).
\end{eqnarray}
The remarkable feature of the TripoSH formalism is to parameterize departures from statistical isotropy regarding total angular momenta, $J$ and $M_{J}$. If the assumption of statistical isotropy breaks, the corresponding expansion coefficients yield the $J\geq 1$ modes, otherwise only the $J=0$ is non-zero. A similar decomposition formalism in two-point statistics, the bipolar spherical harmonic expansion, has been used to search for the breaking of the statistical isotropy assumption of the Universe in the CMB (e.g., see~\citealt{Planck2016A&A...594A..16P,Planck2016A&A...594A..20P} and references therein) and large-scale structure~\citep{Pullen2010JCAP...05..027P,Shiraishi:2016wec,Sugiyama:2017ggb}.

In this work, we assume that the three-point function has arisen through a physical process which is statistically isotropic and parity-symmetric as well as homogeneous. The statistical isotropy induces only the $J=0$ mode, i.e. zero total angular momentum, in the TripoSH formalism, 
resulting in 
\begin{eqnarray}
  \hspace{-0.25cm} B_{\ell_1\ell_2\ell_{12}L}^{J=0,M_J=0}(k_1,k_2) \hspace{-0.25cm} &=& 
    \delta_{\ell_{12}L}^{(K)}(-1)^{\ell_1-\ell_2+L} \nonumber \\
    \hspace{-0.25cm} &\times& \hspace{-0.25cm}
    \sum_{m_1m_2M} \left( \begin{smallmatrix} \ell_1 & \ell_2 & L \\ m_1 & m_2 & M \end{smallmatrix}  \right) B_{\ell_1\ell_2L}^{m_1m_2M}(k_1,k_2), 
\end{eqnarray}
where the statistical isotropy is satisfied through the summation weighted by the Wigner $3$-$j$ symbol over all possible $m_1$, $m_2$, and $M$ with $m_1+m_2+M=0$, and $\delta_{\ell_{12}L}^{(K)}$ is the the Kronecker delta defined such that $\delta_{\ell_{12}L}^{(K)}=1$ if $\ell_{12}=L$, otherwise zero.
Since the parity symmetry restricts allowed multipoles to $\ell_1+\ell_2+L={\rm even}$, it is useful to define the following bispectrum multipoles to simplify the final expressions:
\begin{eqnarray}
	&&\hspace{-0.40cm}B_{\ell_1\ell_2L}(k_1,k_2) \nonumber \\
	&=&
	H_{\ell_1\ell_2L}	\sum_{m_1m_2M}	  \left( \begin{smallmatrix} \ell_1 & \ell_2 & L \\ m_1 & m_2 & M \end{smallmatrix}  \right) 
	B_{\ell_1\ell_2L}^{m_1m_2M}(k_1,k_2), 
	\label{Eq:reduced_ones}
\end{eqnarray}
where $H_{\ell_1\ell_2L}=\left( \begin{smallmatrix} \ell_1 & \ell_2 & L \\ 0 & 0 & 0 \end{smallmatrix}  \right)$ filters even $\ell_1+\ell_2+L$ components.
The bispectrum multipoles defined in the above correspond to the expansion coefficients of the following bispectrum decomposition:
\begin{eqnarray}
	B(\VEC{k}_1,\VEC{k}_2,\hat{n})
	\hspace{-0.25cm}&=&\hspace{-0.7cm}
	\sum_{\ell_1+\ell_2+L={\rm even}} B_{\ell_1\ell_2 L}(k_1,k_2)\, S_{\ell_1\ell_2L}(\hat{k}_1,\hat{k}_2,\hat{n}),
	\label{Eq:TripoSH_J0}
\end{eqnarray}
with the basis function
\begin{eqnarray}
	S_{\ell_1\ell_2L}(\hat{k}_1,\hat{k}_2,\hat{n}) 
   \hspace{-0.25cm} &=& \hspace{-0.25cm}
   \frac{1}{H_{\ell_1\ell_2L}} \sum_{m_1m_2M}  \left( \begin{smallmatrix} \ell_1 & \ell_2 & L \\ m_1 & m_2 & M \end{smallmatrix}  \right) \nonumber \\
    &\times& y_{\ell_1}^{m_1}(\hat{k}_1) y_{\ell_2}^{m_2}(\hat{k}_2) y_L^M(\hat{n}).
    \label{Eq:Slll}
\end{eqnarray} 
The bispectrum multipoles contain all physical information under the three statistical assumptions: homogeneity, isotropy, and parity-symmetry of the Universe.

In the same manner, as the bispectrum, we can expand the three-point function in spherical harmonics,
\begin{eqnarray}
	\zeta(\VEC{r}_1,\VEC{r}_2,\hat{n}) &=&  \sum_{\ell_1\ell_2 L} \sum_{m_1m_2M}\zeta_{\ell_1\ell_2L}^{m_1m_2M}(r_1,r_2) \nonumber \\
	&\times&  y_{\ell_1}^{m_1}(\hat{r}_1) y_{\ell_2}^{m_2}(\hat{r}_2) y_{L}^{M}(\hat{n}).
\end{eqnarray}
Under the assumptions of statistical isotropy and parity-symmetry, we have
\begin{eqnarray}
	\zeta(\VEC{r}_1,\VEC{r}_2,\hat{n})
	\hspace{-0.25cm}&=&\hspace{-0.7cm}
	\sum_{\ell_1+\ell_2+L={\rm even}} \zeta_{\ell_1\ell_2 L}(r_1,r_2)\, S_{\ell_1\ell_2L}(\hat{r}_1,\hat{r}_2,\hat{n}),
	\label{Eq:Zeta_TripoSH_J0}
\end{eqnarray}
and the three-point function multipoles are given by
\begin{eqnarray}
	&&\hspace{-0.4cm}\zeta_{\ell_1\ell_2L}(r_1,r_2) \nonumber \\
	&=&
	H_{\ell_1\ell_2L} 
	 \sum_{m_1m_2M}	  \left( \begin{smallmatrix} \ell_1 & \ell_2 & L \\ m_1 & m_2 & M \end{smallmatrix}  \right) 
	 \zeta_{\ell_1 \ell_2 L}^{m_1m_2M}(r_1,r_2). 
	 \label{Eq:3pcfmultipoles}
\end{eqnarray}
The bispectrum and three-point function multipoles are related to each other according to Hankel transformations:
\begin{eqnarray}
	B_{\ell_1\ell_2L}(k_1,k_2)
	\hspace{-0.25cm}&=&\hspace{-0.25cm} 
	(-i)^{\ell_1+\ell_2}(4\pi)^2 \int dr_1 r_1^2 \int dr_2 r_2^2  \nonumber \\
	&\times& j_{\ell_1}(k_1r_1) j_{\ell_2}(k_2r_2) \zeta_{\ell_1\ell_2L}(r_1,r_2) \nonumber \\
	\zeta_{\ell_1\ell_2L}(r_1,r_2)
	\hspace{-0.25cm}&=&\hspace{-0.25cm} 
	i^{\ell_1+\ell_2}\int \frac{dk_1k_1^2}{2\pi^2} \int \frac{dk_2k_2^2}{2\pi^2} \nonumber \\
	&\times&j_{\ell_1}(r_1k_1)j_{\ell_2}(r_2k_2)B_{\ell_1\ell_2L}(k_1,k_2),
	\label{Eq:HankelTransform}
\end{eqnarray}
where $j_{\ell}$ are the spherical Bessel functions of order $\ell$. While the three-point function multipoles satisfy the reality condition: $\zeta_{\ell_1\ell_2L}^{*}=\zeta_{\ell_1\ell_2L}$, the bispectrum multipoles $B_{\ell_1\ell_2L}$ may be imaginary if $\ell_1+\ell_2={\rm odd}$.

In the context of RSDs, we can show in perturbation theories that the main contribution to the $L$ mode of the bispectrum comes from the terms proportional to $(\hat{n}\cdot\VEC{v})^{L/2}$ (Sugiyama et al. in preparation). Therefore, we expect that the $L$ mode has similar information on the velocity field to the $\ell$ mode of the power spectrum multipoles, expanded in Legendre polynomials. To validate this expectation, we will show in Section~\ref{Sec:CovarianceMatrix} that there is a strong correlation between the bispectrum multipoles and the power spectrum multipoles if $L=\ell$, by measuring the cross-covariance matrix between them from mock catalogues described in Section~\ref{Sec:Data}.

We end this section by discussing allowed combinations of the multipoles, $\ell_1$, $\ell_2$, and $L$. First, since $B(\VEC{k}_1,\VEC{k}_2,\hat{n})=B(\VEC{k}_2,\VEC{k}_1,\hat{n})$, we obtain $B_{\ell_1\ell_2L}(k_1,k_2) = B_{\ell_2\ell_1L}(k_2,k_1)$. Therefore, we can focus only on $\ell_1\geq \ell_2$ without loss of generality. Second, anisotropies due to RSDs are axially symmetric around the LOS direction in the plane-parallel approximation. The axial symmetry then restricts us to $L={\rm even}$. Third, the Wigner $3$-$j$ symbol $\left( \begin{smallmatrix} \ell_1 & \ell_2 & L \\ m_1 & m_2 & M \end{smallmatrix}  \right)$ in equation~(\ref{Eq:reduced_ones}), which appears as a result of the rotational symmetry assumption, satisfies the following selection rule: $|\ell_1-\ell_2| \leq L \leq |\ell_1+\ell_2|$. Finally, with the parity symmetry condition $\ell_1+\ell_2+L=\textrm{even}$, we find $\ell_1+\ell_2=\textrm{even}$ because of $L=\textrm{even}$, resulting in that the bispectrum multipoles should be real: $B_{\ell_1\ell_2L}=B_{\ell_1\ell_2L}^*$. 
Throughout this paper, we focus especially on the $L=0$, $2$, and $4$ modes to clarify the relation to the power spectrum monopole, quadrupole, and hexadecapole. As a demonstration, we write down the first four combinations of $(\ell_1,\ell_2,L)$ for these modes in Table~\ref{Table:allowd_multipoles}. Specifically, in what follows we analyze six bispectrum multipoles $B_{\ell_1\ell_2L}$ for $(\ell_1,\ell_2,L)=(0,0,0)$, $(1,1,0)$, $(2,2,0)$, $(2,0,2)$, $(1,1,2)$, and $(4,0,4)$.

\begin{table}
\centering
\begin{tabular}{cccccc}
\hline\hline
$L$ & \multicolumn{5}{c}{$(\ell_1,\ell_2)$} \\
\hline  
$L=0$ & $(0,0)$ & $(1,1)$ & $(2,2)$ & $(3,3)$ & $\cdots$ \\
$L=2$ & $(2,0)$ & $(1,1)$ & $(3,1)$ & $(2,2)$ & $\cdots$ \\
$L=4$ & $(4,0)$ & $(3,1)$ & $(2,2)$ & $(5,1)$ & $\cdots$ \\
\hline  
\end{tabular}
\caption{
Allowed combinations of the multipoles $(\ell_1,\ell_2,L)$ in $\zeta_{\ell_1\ell_2L}$ and $B_{\ell_1\ell_2L}$,
which are determined by the four conditions: (1) $\ell_1\geq\ell_2$; (2) $L={\rm even}$; (3) $|\ell_1-\ell_2|\leq L\leq |\ell_1+\ell_2|$; and (4) $\ell_1+\ell_2+L={\rm even}$.
As a demonstration, we show here the first four combinations for each of $L=0$, $2$, and $4$.
}
\label{Table:allowd_multipoles}
\end{table}

\section{COMPARISON WITH PREVIOUS WORKS}
\label{Sec:ComparisonWithPreviousWorks}

In the previous section, we have presented a new decomposition formalism which is fairly general and independent of the choice of the coordinate system. In this section, we will show how our formalism is related to other coordinate choices used in previous work.

\subsection{The $L=0$ mode, i.e., monopole}

In the absence of both the RSD and AP effects, the bispectrum (the three-point function) is a function of $\VEC{k}_1$ and $\VEC{k}_2$ ($\VEC{r}_1$ and $\VEC{r}_2$), and hence, it is common to expand it in Legendre polynomials ${\cal L}_{\ell}$~\citep{Szapudi:2004gg,Pan:2005ym,Slepian:2015qza,Slepian:2016qwa}:
\begin{eqnarray}
	B(\VEC{k}_1,\VEC{k}_2) &=& \sum_{\ell}\, B_{\ell}(k_1,k_2)\, {\cal L}_{\ell}(\hat{k}_1\cdot\hat{k}_2) \nonumber \\
	\zeta(\VEC{r}_1,\VEC{r}_2) &=& \sum_{\ell}\, \zeta_{\ell}(r_1,r_2)\, {\cal L}_{\ell}(\hat{r}_1\cdot\hat{r}_2).
\end{eqnarray}
This decomposition is possible even in redshift space after averaging over the LOS: $B(\VEC{k}_1,\VEC{k}_2) = \int \frac{d^2\hat{n}}{4\pi} B(\VEC{k}_1,\VEC{k}_2,\hat{n})$. The corresponding expansion coefficients $B_{\ell}$ (hereafter, the Legendre coefficients) are then given by
\begin{eqnarray}
	B_{\ell}(k_1,k_2)
	&=& 
	(2\ell+1)\int \frac{d^2\hat{k}_1}{4\pi}\int \frac{d^2\hat{k}_2}{4\pi}\int \frac{d^2\hat{n}}{4\pi}  \nonumber \\
	&\times&
	{\cal L}_{\ell}(\hat{k}_1\cdot\hat{k}_2)\, B(\VEC{k}_1,\VEC{k}_2,\hat{n}),
	\label{Eq:LegendreCoefficients}
\end{eqnarray}
and $\zeta_{\ell}(r_1,r_2)$ can be computed using a Hankel transform~\cite[equation~3 in][]{Szapudi:2004gg}:
\begin{eqnarray}
	\zeta_{\ell}(r_1,r_2)\hspace{-0.25cm}&=& \hspace{-0.25cm}
	(-1)^{\ell}\int \frac{dk_1k_1^2}{2\pi^2} \int \frac{dk_2k_2^2}{2\pi^2} \nonumber \\
	&\times&j_{\ell}(r_1k_1)j_{\ell}(r_2k_2)B_{\ell}(k_1,k_2).
	\label{Eq:Bell2zetaell}
\end{eqnarray}

The $L=0$ mode of $\zeta_{\ell_1\ell_2L}$ and $B_{\ell_1\ell_2L}$ reproduces the Legendre coefficients $\zeta_{\ell}$ and $B_{\ell}$. Using the relation between spherical harmonics and Legendre polynomials (equation~\ref{Eq:LYlm}), equation~(\ref{Eq:reduced_ones}) for $L=0$ leads to
\begin{eqnarray}
	\zeta_{\ell_1\ell_2 L=0}(r_1,r_2) &=& \delta^{\rm (K)}_{\ell_1\ell_2}\,\zeta_{\ell_1}(r_1,r_2) \nonumber \\
	B_{\ell_1\ell_2 L=0}(k_1,k_2) &=& \delta^{\rm (K)}_{\ell_1\ell_2}\,B_{\ell_1}(k_1,k_2).
	\label{Eq:Legendre}
\end{eqnarray}
Equation~(\ref{Eq:HankelTransform}) for $L=0$ reduces to equation~(\ref{Eq:Bell2zetaell}). Thus, our formalism can be regarded as a generalized one of the Legendre coefficients, and the $L\geq 2$ modes will provide additional information in terms of anisotropic signals induced by the RSD or AP effects.

\subsection{Coordinate systems}
 
\begin{figure}
	\includegraphics[width=\columnwidth]{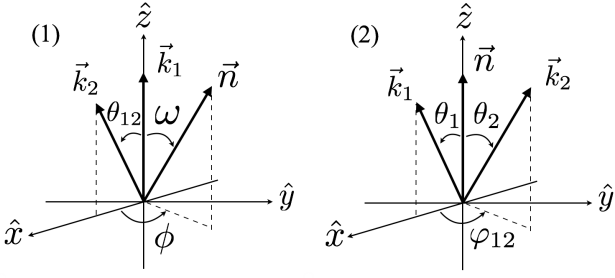}
	\caption{
		The configuration of the variables used to compute the bispectrum multipoles.
		The left and right coordinate systems choose $\hat{k}_1$ and $\hat{n}$ as the $z$-axis, respectively.
	}
	\label{fig:coordinates}
\end{figure}

While the bispectrum and three-point function multipoles shown in Section~\ref{Sec:NewDecompositionFormalsim} are independent of the choice of coordinate systems, it is convenient to choose specific coordinate axes when theoretically modeling them. In this section, we present two choices of coordinates: (1) $\hat{k}_1$ is chosen as the $\hat{z}$-axis, which is frequently used for the bispectrum in the literature~\citep[e.g.,][]{Scoccimarro:1999ed}; and (2) the LOS direction $\hat{n}$ is taken as the $\hat{z}$-axis, which is recently proposed by~\cite{Slepian:2017lpm} to characterize RSDs in the three-point function. The configuration of the variables in these coordinate systems is shown in Fig.~\ref{fig:coordinates}. We present how to calculate the bispectrum multipoles (equation~\ref{Eq:reduced_ones}) in these specific coordinates; the three-point function multipoles can then be computed through the Hankel transform in equation~(\ref{Eq:HankelTransform}).
In Section~\ref{Sec:SurveyWindowFunctions}, we adopt Scoccimarro's coordinate system to compute the theoretical bispectrum multipoles using perturbation theory.

\subsubsection{Case 1: $\hat{k}_1=\hat{z}$}
\label{Sec:KZ}

Following~\cite{Scoccimarro:1999ed}, we choose $\hat{k}_1$ as the $z$-axis without loss of generality and adopt the following coordinate system: 
\begin{eqnarray}
	\VEC{k}_1 &=& \{0,\,0,\,k_1\} \nonumber \\
	\VEC{k}_2 &=&  \{k_2\sin \theta_{12},\, 0,\, k_2\cos\theta_{12}\} \nonumber \\
	\hat{n}&=& \{\sin\omega \cos \phi,\, \sin\omega\sin\phi,\, \cos\omega\}.
\end{eqnarray}
The bispectrum in redshift space is then characterized by the five parameters: $B = B(k_1,k_2,\theta_{12},\omega,\phi)$. The three parameters $k_1$, $k_2$, and $\theta_{12}$ define the shape of the triangle; the remaining two angles $\omega$ and $\phi$ characterize the orientation of the triangle with respect to the LOS direction. The spherical harmonic functions with $\hat{k}_1$, $\hat{k}_2$, and $\hat{n}$ then become $y_{\ell_1}^{m_1*}(\hat{z})=\delta^{\rm (K)}_{0 m_1}$, $y_{\ell_2}^{m_2*}(\hat{k}_2)=y_{\ell_2}^{m_2*}(\theta_{12}, 0)$, and $y^{M*}_{L}(\hat{n}) = y^{M*}_{L}(\omega, \phi)$, respectively. Substituting these into equations~(\ref{Eq:SH_coefficients}) and (\ref{Eq:reduced_ones}) leads to
\begin{eqnarray}
	B_{\ell_1\ell_2L}(k_1,k_2) 
	\hspace{-0.25cm}&=&\hspace{-0.25cm}
	N_{\ell_1\ell_2L}\, H_{\ell_1\ell_2L}\,
	\int \frac{d \cos \omega d\phi}{4\pi}\int \frac{d \cos \theta_{12}}{2}\nonumber \\
	\hspace{-0.25cm}&\times&\hspace{-0.25cm}
	\left[\sum_{M} \left( \begin{smallmatrix} \ell_1 & \ell_2 & L \\ 0 & -M & M \end{smallmatrix}  \right) 
	y^{-M*}_{\ell_2}(\theta_{12}, 0)y^{M*}_{L}(\omega, \phi) \right]
	\nonumber \\
	\hspace{-0.25cm}&\times&\hspace{-0.25cm}
	B(k_1,k_2, \theta_{12},\omega,\phi).
	\label{Eq:B1}
\end{eqnarray}

To more clearly show the relation to the previous work, we decompose $B(k_1,k_2,\theta_{12},\omega,\phi)$ into spherical harmonics~\cite[equation~19 in][]{Scoccimarro:1999ed}:
\begin{eqnarray}
	B(k_1,k_2, \theta_{12},\omega,\phi)=\sum_{LM}B_{LM}(k_1,k_2, \theta_{12}) Y_{L}^M(\omega,\phi).
\end{eqnarray}
Then, the above bispectrum multipoles $B_{LM}$ are related to ours $B_{\ell_1\ell_2L}$ as follows:
\begin{eqnarray}
	B_{\ell_1\ell_2L}(k_1,k_2) 
	\hspace{-0.25cm}&=&\hspace{-0.25cm}
	\frac{N_{\ell_1\ell_2L} H_{\ell_1\ell_2L}}{\sqrt{(4\pi)(2L+1)}}\int \frac{d \cos \theta_{12}}{2}\nonumber \\
	\hspace{-0.25cm}&\times&\hspace{-0.25cm}
	\left[\sum_{M} \left( \begin{smallmatrix} \ell_1 & \ell_2 & L \\ 0 & -M & M \end{smallmatrix}  \right) 
	y^{-M*}_{\ell_2}(\cos\theta_{12}, 0) \right]
	\nonumber \\
	\hspace{-0.25cm}&\times&\hspace{-0.25cm}
	B_{LM}(k_1,k_2, \theta_{12}).
	\label{Eq:B2}
\end{eqnarray}

\subsubsection{Case 2: $\hat{n}=\hat{z}$}
\label{Sec:NZ}

We choose $\hat{n}$ as the $z$-axis and adopt the following coordinate system: 
\begin{eqnarray}
	\VEC{k}_1 &=& \{k_1\sin\theta_1,\, 0,\, k_1\cos\theta_1\} \nonumber \\
	\VEC{k}_2 &=& \{k_2\sin\theta_2 \cos \varphi_{12},\, k_2\sin \theta_2\sin\varphi_{12},\, k_2\cos\theta_2\} \nonumber \\
	\hat{n} &=& \{0,\,0,\,1\}.
\end{eqnarray}
In the same manner as Section~\ref{Sec:KZ}, we obtain
\begin{eqnarray}
	B_{\ell_1\ell_2L}(k_1,k_2) 
	\hspace{-0.25cm}&=&\hspace{-0.25cm}
	N_{\ell_1\ell_2L} H_{\ell_1\ell_2L}
	\int \frac{d\cos \theta_1}{2}\int \frac{d\cos \theta_2 d\varphi_{12}}{4\pi} 
	\nonumber \\
	\hspace{-0.25cm}&\times&\hspace{-0.25cm}
	\left[\sum_{m}\left( \begin{smallmatrix} \ell_1 & \ell_2 & L \\ m & -m & 0 \end{smallmatrix}  \right) 
	y_{\ell_1}^{m*}(\theta_1,0)
	y_{\ell_2}^{-m*}(\theta_2,\varphi_{12})  \right]\nonumber \\
	\hspace{-0.25cm}&\times&\hspace{-0.25cm}
	B(k_1,k_2,\theta_1,\theta_2, \varphi_{12}).
	\label{Eq:B3}
\end{eqnarray}

\citet{Slepian:2017lpm} proposed to decompose the three-point function into a product of two spherical harmonics:
$\zeta(\VEC{r}_1,\VEC{r}_2,\hat{n}=\hat{z}) = \sum_{\ell_1\ell_2m} \zeta_{\ell_1\ell_2}^m(r_1,r_2) y_{\ell_1}^m(\hat{r}_1) y_{\ell_2}^{m*}(\hat{r}_2)$.
The Fourier transform of the expansion coefficients $\zeta_{\ell_1\ell_2}^m$ is given by
\begin{eqnarray}
		B_{\ell_1\ell_2}^m(k_1,k_2) 
	\hspace{-0.25cm}&=&\hspace{-0.25cm}
	(2\ell_1+1)(2\ell_2+1)
	\int \frac{d\cos \theta_1}{2}\int \frac{d\cos \theta_2 d\varphi_{12}}{4\pi} 
	\nonumber \\
	\hspace{-0.25cm}&\times&\hspace{-0.25cm}
	y_{\ell_1}^{m*}(\theta_1,0)
	y_{\ell_2}^{m}(\theta_2,\varphi_{12})  \nonumber \\
	\hspace{-0.25cm}&\times&\hspace{-0.25cm}
	B(k_1,k_2,\theta_1,\theta_2, \varphi_{12}).
\end{eqnarray}
Equation~(\ref{Eq:B3}) can be then represented as
\begin{eqnarray}
	&&\hspace{-0.4cm}B_{\ell_1\ell_2L}(k_1,k_2) \nonumber \\
	\hspace{-0.25cm}&=&\hspace{-0.25cm}
	(2L+1) H_{\ell_1\ell_2L}
    \sum_{m}\left( \begin{smallmatrix} \ell_1 & \ell_2 & L \\ m & -m & 0 \end{smallmatrix}  \right) (-1)^{m} B_{\ell_1\ell_2}^m(k_1,k_2).
\end{eqnarray}
The three-point function multipoles presented in this paper are
related to $\zeta_{\ell_1\ell_2}^m$ as follows:
\begin{eqnarray}
	&&\hspace{-0.4cm}\zeta_{\ell_1\ell_2L}(r_1,r_2) \nonumber \\
	\hspace{-0.25cm}&=&\hspace{-0.25cm}
	(2L+1) H_{\ell_1\ell_2L}
    \sum_{m}\left( \begin{smallmatrix} \ell_1 & \ell_2 & L \\ m & -m & 0 \end{smallmatrix}  \right) (-1)^{m} \zeta_{\ell_1\ell_2}^m(r_1,r_2).
\end{eqnarray}

\section{ESTIMATORS}
\label{Sec:Estimator}

In this section, we present the estimators to measure the bispectrum and three-point function multipoles, $B_{\ell_1\ell_2L}$ and $\zeta_{\ell_1\ell_2L}$, using the FFT algorithm. The algorithm to measure the three-point statistics using FFT was first proposed by~\cite{Scoccimarro:2015bla} for the bispectrum, and by~\cite{Slepian:2016qwa} for the three-point function. We extend their estimators to our decomposition formalism. Section~\ref{Sec:number_density} describes how to measure the number density of galaxies weighted by spherical harmonics with the LOS direction $y_{L}^{M*}(\hat{x})$. Using the $y_{L}^{M*}$-weighted density field, we build the estimators of $B_{\ell_1\ell_2L}$ and $\zeta_{\ell_1\ell_2L}$ in Sections~\ref{Sec:bispectrum} and~\ref{Sec:threePT_func}. While these estimators can be applied to an observed galaxy catalog, in Section~\ref{Sec:GlobalLos} we show that our estimator can be straightforwardly applied to $N$-body simulations in which the global-plane parallel approximation is often adopted.

Measuring the bispectrum and three-point function from galaxy data, we use the Cartesian coordinates such that $\VEC{x}=$ $\chi(z)\{\cos\delta_{\rm g}\cos\alpha_{\rm g}, \cos\delta_{\rm g}\sin\alpha_{\rm g},\sin\delta_{\rm g} \}$, where $\chi$ is the comoving radial distance to galaxies, $\alpha_{\rm g}$ and $\delta_{\rm g}$ are respectively the right ascension and declination angles of galaxies. To be consistent with such data coordinate system, we choose the north pole as our $z$-axis for our decomposition. Again, the resulting bispectrum multipoles in our formalism do not depend on this axis choice.

\subsection{Number density of galaxies}
\label{Sec:number_density}

The number density of galaxies is given by
\begin{eqnarray}
	  n(\VEC{x}) =  \sum_{i}^{N_{\rm gal}} w(\VEC{x}_i) \delta_{\rm D}\left( \VEC{x} - \VEC{x}_i \right),
\end{eqnarray}
where $\VEC{x}_i$ represents the observed position of galaxy $i$, $N_{\rm gal}$ denotes the total number of observed galaxies, and the weight function $w(\VEC{x})$ may include systematic weights~\cite[e.g.,][]{Reid:2015gra} and the FKP weight~\citep{Feldman:1993ky} (for details, see Section~\ref{Sec:Prescription}). We estimate the mean number density $\bar{n}(\VEC{x})$ from a synthetic random catalogue,
\begin{eqnarray}
	\bar{n}(\VEC{x}) = \alpha \sum_{i}^{N_{\rm ran}} w(\VEC{x}_i) \delta_{\rm D}\left( \VEC{x} - \VEC{x}_i \right),
	\label{Eq:mean_n}
\end{eqnarray}
where $N_{\rm ran}$ is the total number of objects in the random catalogue, and $\alpha = \sum_{i}^{N_{\rm gal}}w_i/\sum_{i}^{N_{\rm ran}} w_i$. By subtracting $\bar{n}(\VEC{x})$ from $n(\VEC{x})$, we obtain the density fluctuation:
\begin{eqnarray}
	\delta n(\VEC{x}) = n(\VEC{x}) - \bar{n}(\VEC{x}).
	\label{Eq:dn}
\end{eqnarray}
We then define the $y_{L}^{M*}$-weighted density fluctuation as
~\citep{Sugiyama:2017ggb,Hand:2017irw}
\begin{eqnarray}
	\delta n_{L}^M(\VEC{x})
	 \hspace{-0.25cm}&\equiv&  \hspace{-0.25cm}
	 y_{L}^{M*}(\hat{x})\, \delta n(\VEC{x})\nonumber \\
	\hspace{-0.25cm}&=&  \hspace{-0.28cm}
	\left( \sum_{i}^{N_{\rm gal}} - \alpha\sum_{i}^{N_{\rm ran}}  \right) 
	y_{L}^{M*}(\hat{x}_i) w(\VEC{x}_i) \delta_{\rm D}\left( \VEC{x} - \VEC{x}_i \right). \nonumber \\
	\label{Eq:nlm}
\end{eqnarray}
We note here that the leading order of $\delta n_{L}^M$ is the same as the normal density fluctuation: $\delta n(\VEC{x})=\delta n_{0}^0(\VEC{x})$.

\subsection{Bispectrum}
\label{Sec:bispectrum}

Since the LOS direction is not globally but locally determined by galaxy positions, we apply the local plane-parallel approximation $\hat{x}_1\approx \hat{x}_2\approx \hat{x}_3$ and choose $\hat{x}_3$ as the LOS direction in the main text (see Section~\ref{Sec:NewDecompositionFormalsim} and Appendix~\ref{Ap:LOS}). Under this approximation, we present the estimator of the bispectrum multipoles as follows:
\begin{eqnarray}
	\widehat{B}_{\ell_1\ell_2L}(k_1,k_2)
	\hspace{-0.25cm} &=&\hspace{-0.25cm}
	 H_{\ell_1\ell_2L}\, \sum_{m_1m_2M}	  \left( \begin{smallmatrix} \ell_1 & \ell_2 & L \\ m_1 & m_2 & M \end{smallmatrix}  \right)  \nonumber \\
	\hspace{-0.25cm} &\times&\hspace{-0.25cm}
	\frac{N_{\ell_1\ell_2L}}{I}\, \int \frac{d^2\hat{k}_1}{4\pi} y_{\ell_1}^{m_1*}(\hat{k}_1)
	  \int \frac{d^2\hat{k}_2}{4\pi} y_{\ell_2}^{m_2*}(\hat{k}_2) \nonumber \\
	\hspace{-0.25cm} &\times&\hspace{-0.25cm}
	  \int \frac{d^3k_3}{(2\pi)^3} (2\pi)^3 \delta_{\rm D}\left( \VEC{k}_1+ \VEC{k}_2+ \VEC{k}_3 \right)  \nonumber \\
	\hspace{-0.25cm} &\times&\hspace{-0.25cm}
	  \delta n(\VEC{k}_1)\,  \delta n(\VEC{k}_2)\,  \delta n_{L}^M(\VEC{k}_3),
	  \label{Eq:bi_estimator}
\end{eqnarray}
where the triangle condition $\VEC{k}_1+\VEC{k}_2+\VEC{k}_3=0$ is satisfied through the delta function, and $\delta n_{L}^M(\VEC{k})$ is the Fourier transform of the $y_{L}^{M*}$-weighted density fluctuation (equation~\ref{Eq:nlm}): $\delta n_{L}^M(\VEC{k}) = \int d^3x \mathrm{e}^{-\mathrm{i}\VEC{k}\cdot\VEC{x}} \delta n_{L}^{M}(\VEC{x})$ and $\delta n(\VEC{k}) = \delta n_0^0(\VEC{k})$. As mentioned in Section~\ref{Sec:NewDecompositionFormalsim}, the $L=0$ mode reduces to the estimator of the Legendre coefficients of the bispectrum:
\begin{eqnarray}
	\widehat{B}_{\ell}(k_1,k_2) &=& \widehat{B}_{\ell \ell L=0}(k_1,k_2) \nonumber \\
	\hspace{-0.25cm} &=&\hspace{-0.25cm}
	\frac{2\ell+1}{I}\, \int \frac{d^2\hat{k}_1}{4\pi}  \int \frac{d^2\hat{k}_2}{4\pi} {\cal L}_{\ell}(\hat{k}_1\cdot\hat{k}_2) \nonumber \\
	\hspace{-0.25cm} &\times&\hspace{-0.25cm}
	  \int \frac{d^3k_3}{(2\pi)^3} (2\pi)^3 \delta_{\rm D}\left( \VEC{k}_1+ \VEC{k}_2+ \VEC{k}_3 \right)  \nonumber \\
	\hspace{-0.25cm} &\times&\hspace{-0.25cm}
	  \delta n(\VEC{k}_1)\,  \delta n(\VEC{k}_2)\,  \delta n(\VEC{k}_3).
\end{eqnarray}

The normalization $I$ in equation.~(\ref{Eq:bi_estimator}) is given by
\begin{eqnarray}
	I = \int d^3x\, \bar{n}^3(\VEC{x}).
	\label{Eq:normalization}
\end{eqnarray}
This normalization depends on a grid-cell resolution to compute the mean density $\bar{n}(\VEC{x})$, and hence, it is difficult to make $I$ numerically converge to a certain value in a large survey volume. However, the value of $I$ does not affect the final results, 
because three-point window functions, which are defined in Section~\ref{Sec:SurveyWindowFunctions} to estimate survey geometry effects, have the same normalization, and the theoretical model of the bispectrum is represented as a convolution of the theory and the window function. Namely, 
the normalization $I$ cancels out when comparing the theory including the survey window effect with the measurements.

The angular integrations in equation~(\ref{Eq:bi_estimator}) represent integration over a spherical shell in Fourier space centered at each bin $k=|\VEC{k}|$,
\begin{eqnarray}
	  \int \frac{d^2\hat{k}}{4\pi}
	  &=&  \frac{(2\pi)^3}{4\pi k^2}\int \frac{d^3k'}{(2\pi)^3} \delta_{\rm D}\left( |\VEC{k}| - |\VEC{k}'| \right) \nonumber \\
	  &=& 
	  \frac{1}{N_{\rm mode}(k)}\sum_{k-\Delta k/2<k<k+\Delta/2}
	  \label{Eq:angular_integral}
\end{eqnarray}
where $N_{\rm mode}(k)$ is the number of Fourier modes in each $k$-bin and $\Delta k$ is the bin width. 

The FFT algorithm to compute $\delta n_{L}^M(\VEC{k})$ requires the interpolation of functions on a regular grid in position space. The Fourier transform of the $y_{L}^{M*}$-weighted density fluctuation measured by FFTs, $\delta n_{L}^{M}(\VEC{k})|_{\rm FFT}$, includes the effect of the mass assignment function $W_{\rm mass}(\VEC{k})$~\citep{Jing:2004fq}. Although there are some efforts to reduce the aliasing effect~\cite[e.g.,][]{Sefusatti:2015aex}, we adopt in this paper the simplest method to correct such effects: we divide $\delta n_{L}^M(\VEC{k})|_{\rm FFT}$ by $W_{\rm mass}(\VEC{k})$, namely $\delta n_{L}^M(\VEC{k}) = \delta n_{L}^M(\VEC{k})|_{\rm FFT}/W_{\rm mass}(\VEC{k})$. The mass assignment window function is given by
\begin{eqnarray}
	  W_{\rm mass}(\VEC{k}) = \prod_{i=x,y,z}\left[ {\rm sinc}\left( \frac{\pi k_i}{2 k_{{\rm N}}} \right) \right]^{p},
\end{eqnarray}
where ${\rm sinc}(x) = \sin(x)/x$, and $k_{{\rm N}}=\pi/H_{\rm G}$ is the Nyquist frequency with the grid spacing $H_{\rm G}$ on an axis. The indexes $p=1$, $p=2$, and $p=3$ correspond to the nearest grid point (NGP), cloud-in-cell (CIC), and triangular-shaped cloud (TSC) assignment functions, respectively~\citep{HockneyEastwood1981}. Appendix~\ref{Ap:MassAssignmentFunction} investigates the differences between the three mass assignment functions.

The integrals in equation~(\ref{Eq:bi_estimator}) do not decouple into a product of Fourier transforms due to the delta function $\delta_{\rm D}\left( \VEC{k}_1+\VEC{k}_2+\VEC{k}_3 \right)$. When discretizing the integrals, they become double sums in the number of grid-cells, leading to an $N_{\rm grid}^2$ process, where $N_{\rm grid}$ is the number of FFT grid-cells. To compute these integrals using FFTs, we insert the relation
\begin{eqnarray}
	   (2\pi)^3 \delta_{\rm D}\left( \VEC{k}_1+ \VEC{k}_2+ \VEC{k}_3 \right)
	   =  \int d^3x\, \mathrm{e}^{\mathrm{i}\VEC{x}\cdot(\VEC{k}_1+\VEC{k}_2+\VEC{k}_3)}
\end{eqnarray}
in equation.~(\ref{Eq:bi_estimator}), and then derive
\begin{eqnarray}
	\hspace{-0.35cm}\widehat{B}_{\ell_1\ell_2L}(k_1,k_2)
	\hspace{-0.25cm}&=& \hspace{-0.25cm}
	H_{\ell_1\ell_2L}\, \sum_{m_1m_2M}	  \left( \begin{smallmatrix} \ell_1 & \ell_2 & L \\ m_1 & m_2 & M \end{smallmatrix}  \right)    \nonumber \\
	\hspace{-0.25cm}  &\times&\hspace{-0.25cm}
	\frac{N_{\ell_1\ell_2L}}{I}\,  \int d^3x\, F_{\ell_1}^{\, m_1}(\VEC{x};k_1)\,  F_{\ell_2}^{\, m_2}(\VEC{x};k_2)\,  G_{L}^{M}(\VEC{x}), \nonumber \\
	  \label{Eq:bi_estimator_FFTs}
\end{eqnarray}
where
\begin{eqnarray}
	F_{\ell}^{\, m}(\VEC{x};k) &=&  \int \frac{d^2\hat{k}}{4\pi} \,
	\mathrm{e}^{\mathrm{i}\VEC{k}\cdot\VEC{x}} \, y_{\ell}^{m*}(\hat{k})\, \frac{\delta n|_{\rm FFT}(\VEC{k})}{W_{\rm mass}(\VEC{k})} \nonumber \\
	G_{L}^{\, M}(\VEC{x}) &=&  \int \frac{d^3k}{(2\pi)^3} \,
	\mathrm{e}^{\mathrm{i}\VEC{k}\cdot\VEC{x}} \, \frac{\delta n_{L}^M|_{\rm FFT}(\VEC{k})}{W_{\rm mass}(\VEC{k})}.
	\label{Eq:nlm_bi}
\end{eqnarray}
Since the angular integration in $F_{\ell}^m$ can be represented as a three-dimensional integral with a delta function through equation~(\ref{Eq:angular_integral}), it can be computed using inverse FFTs. Thus, this estimator requires only FFT, i.e. ${\cal O}(N_{\rm grid} \ln N_{\rm grid})$, processes~\citep[see also][]{Scoccimarro:2015bla}. 

Once we have measured the bispectrum multipoles from galaxies using its estimator (equation~\ref{Eq:bi_estimator}), it is necessary to subtract from that the shot noise terms, which arise from a sampling of a catalogue with a finite number of galaxies. To estimate the shot noise in the bispectrum, we substitute equation~(\ref{Eq:nlm}) into equation~(\ref{Eq:bi_estimator}) and write down the bispectrum estimator in the particle description:
\begin{eqnarray}
	&& \hspace{-0.25cm}\widehat{B}_{\ell_1\ell_2L}(k_1,k_2) \nonumber \\
	\hspace{-0.25cm}&=&\hspace{-0.25cm}
	 H_{\ell_1\ell_2L}\, \sum_{m_1m_2M}	  \left( \begin{smallmatrix} \ell_1 & \ell_2 & L \\ m_1 & m_2 & M \end{smallmatrix}  \right)  \nonumber \\
	\hspace{-0.25cm} &\times&\hspace{-0.25cm}
	\frac{N_{\ell_1\ell_2L}}{I}\,  \int \frac{d^2\hat{k}_1}{4\pi} y_{\ell_1}^{m_1*}(\hat{k}_1)
	 \int \frac{d^2\hat{k}_2}{4\pi} y_{\ell_2}^{m_2*}(\hat{k}_2) \nonumber \\
	\hspace{-0.25cm}  &\times& \hspace{-0.25cm}
	  \int \frac{d^3k_3}{(2\pi)^3} (2\pi)^3 \delta_{\rm D}\left( \VEC{k}_1+ \VEC{k}_2+ \VEC{k}_3 \right)  \nonumber \\
	\hspace{-0.25cm}  &\times& \hspace{-0.25cm}
	  \left(    \sum_{i}^{N_{\rm gal}} - \alpha \sum_i^{N_{\rm ran}} \right) 
	  \left(    \sum_{j}^{N_{\rm gal}} - \alpha \sum_j^{N_{\rm ran}} \right) 
	  \left(    \sum_{k}^{N_{\rm gal}} - \alpha \sum_k^{N_{\rm ran}} \right) 
	  \nonumber \\
	\hspace{-0.25cm}  &\times& \hspace{-0.25cm}
	w(\VEC{x}_i) w(\VEC{x}_j) w(\VEC{x}_k) y_{L}^{M*}(\hat{x}_k)
	\mathrm{e}^{-\mathrm{i}\VEC{k}_1\cdot\VEC{x}_i}
	\mathrm{e}^{-\mathrm{i}\VEC{k}_2\cdot\VEC{x}_j}
	\mathrm{e}^{-\mathrm{i}\VEC{k}_3\cdot\VEC{x}_k}, \nonumber \\
\end{eqnarray}
where $\VEC{x}_i$, $\VEC{x}_j$, and $\VEC{x}_k$ denote the positions of galaxies $i$, $j$, and $k$, respectively. Then, there are four cases to contribute to the shot noise: (1) $i=j=k$; (2) $i=k\neq j$; (3) $j=k\neq i$; and (4) $i=j\neq k$. The total shot noise is thus
\begin{eqnarray}
	S_{\ell_1\ell_2L}(k_1,k_2)
	\hspace{-0.25cm}&=&\hspace{-0.25cm}
	S_{\ell_1\ell_2L}\big|_{i=j=k} + S_{\ell_1\ell_2L}\big|_{i=j\neq k} \nonumber \\
	\hspace{-0.25cm}&+&\hspace{-0.25cm}
	S_{\ell_1\ell_2L}\big|_{i=k\neq j} + S_{\ell_1\ell_2L}\big|_{j=k\neq i}.
	\label{Eq:shotnoise}
\end{eqnarray}
Each term of the right-hand side in the above expression is given by
\begin{eqnarray}
	S_{\ell_1\ell_2L}\big|_{i=j=k}
	\hspace{-0.25cm}&=&\hspace{-0.25cm}
	\delta_{\ell_10}^{(\rm K)}\delta_{\ell_20}^{(\rm K)}\delta_{L0}^{(\rm K)}\left( 1/I  \right)\bar{S}_{L=0}^{M=0}\nonumber\\
	S_{\ell_1\ell_2L}|_{i\neq j=k}(k_1)
	\hspace{-0.25cm}&=&\hspace{-0.25cm}
	\delta^{(\rm K)}_{\ell_1 L}\delta^{(\rm K)}_{\ell_2 0}
	 \frac{2L+1}{I} \int \frac{d^2\hat{k}_1}{4\pi} y_{L}^{M}(\hat{k}_1) \nonumber \\
	\hspace{-0.25cm} &\times& \hspace{-0.25cm} \left[ \delta n(\VEC{k}_1) N_L^{M*}(\VEC{k}_1) - \frac{\bar{S}_L^M\,C_{\rm shot}(\VEC{k}_1)}{W_{\rm mass}^2(\VEC{k}_1)} \right] \nonumber \\
	S_{\ell_1\ell_2L}\big|_{i=k\neq j}(k_2)
	\hspace{-0.25cm}&=&\hspace{-0.25cm}
	\delta^{(\rm K)}_{\ell_2L}\delta^{(\rm K)}_{\ell_10} \frac{2L+1}{I} \int \frac{d^2\hat{k}_2}{4\pi} y_{L}^{M}(\hat{k}_2)\nonumber \\
	\hspace{-0.25cm}&\times&\hspace{-0.25cm}\left[  \delta n(\VEC{k}_2) N_L^{M*}(\VEC{k}_2) -  \frac{\bar{S}_L^M\,C_{\rm shot}(\VEC{k}_2)}{W_{\rm mass}^2(\VEC{k}_2)} \right]
	  \nonumber \\
	  S_{\ell_1\ell_2L}\big|_{i=j\neq k}(k_1,k_2) 
	\hspace{-0.25cm}&=&\hspace{-0.25cm}
	 H_{\ell_1\ell_2L}\, \sum_{m_1m_2M}	  \left( \begin{smallmatrix} \ell_1 & \ell_2 & L \\ m_1 & m_2 & M \end{smallmatrix}  \right)  \nonumber \\
	\hspace{-0.25cm} &\times&\hspace{-0.25cm}
	\frac{N_{\ell_1\ell_2L}}{I}i^{\ell_1+\ell_2} 
	  \int d^3x j_{\ell_1}(k_1x)j_{\ell_2}(k_2x) \nonumber \\
	\hspace{-0.25cm}&\times&\hspace{-0.25cm}
	  y_{\ell_1}^{m_1*}(\hat{x})  y_{\ell_2}^{m_2*}(\hat{x}) 
	  \int \frac{d^3k_3}{(2\pi)^3}\mathrm{e}^{\mathrm{i}\VEC{k}_3\cdot\VEC{x}}  \nonumber \\
	  \hspace{-0.25cm}&\times&\hspace{-0.25cm}
	  \left[  \delta n_L^M(\VEC{k}_3) N_0^{0*}(\VEC{k}_3)  - \frac{\bar{S}_L^M \,C_{\rm shot}(\VEC{k}_3)}{W_{\rm mass}^2(\VEC{k}_3)}\right]\nonumber \\
      \label{Eq:ShotNoise}
\end{eqnarray}
where
\begin{eqnarray}
	\bar{S}_L^M \hspace{-0.25cm}&=&\hspace{-0.25cm} \left(    \sum_{i}^{N_{\rm gal}} - \alpha^3 \sum_i^{N_{\rm ran}} \right) [w(\VEC{x}_i)]^3\, y_{L}^{M*}(\hat{x}_i) \nonumber \\
	N_L^M(\VEC{k}) \hspace{-0.25cm}&=& \hspace{-0.25cm}
	 \left(    \sum_{i}^{N_{\rm gal}} + \alpha^2 \sum_i^{N_{\rm ran}} \right) 
	 \left[ w(\VEC{x}_i) \right]^2 y_{L}^{M}(\hat{x}_i) \mathrm{e}^{-\mathrm{i}\VEC{k}\cdot\VEC{x}_i}.
\end{eqnarray}
The function $C_{\rm shot}(\VEC{k})$ is an analytical scale-dependent function to correct for the mass alignment effect on the shot noise~\cite[see equation~20 in][]{Jing:2004fq}. We expect that $S_{\ell_1\ell_2L}|_{i=j=k}$ should also depend on scales when taking into account the mass alignment effect and that we can derive its analytical expression similar to the power spectrum case. However, we leave more carefully investigations of these corrections for future work. 

\subsection{Three-point function}
\label{Sec:threePT_func}

Now we move onto the derivation of the FFT-based estimator of the three-point function multipoles. Since the three-point function multipoles are related to the bispectrum multipoles through the Hankel transform, substituting equation~(\ref{Eq:bi_estimator}) into equation~(\ref{Eq:HankelTransform}) leads to
\begin{eqnarray}
	\widehat{\zeta}_{\ell_1\ell_2L}(r_1,r_2)
	\hspace{-0.25cm}&=&\hspace{-0.25cm}
	 H_{\ell_1\ell_2L}\, \sum_{m_1m_2M}	  \left( \begin{smallmatrix} \ell_1 & \ell_2 & L \\ m_1 & m_2 & M \end{smallmatrix}  \right)  \nonumber \\
	\hspace{-0.25cm} &\times&\hspace{-0.25cm}
	\frac{N_{\ell_1\ell_2L}}{I}\int \frac{d^2\hat{r}_1}{4\pi}y_{\ell_1}^{m_1*}(\hat{r}_1)\int \frac{d^2\hat{r}_2}{4\pi}y_{\ell_2}^{m_2*}(\hat{r}_2)
	\nonumber \\
	\hspace{-0.25cm}&\times&\hspace{-0.25cm} 
	\int d^3x_1\int d^3x_2\int d^3x_3  \nonumber \\
	\hspace{-0.25cm}&\times&\hspace{-0.25cm} 
	\delta_{\rm D}\left( \VEC{r}_1-\VEC{x}_{13}\right) \delta_{\rm D}\left( \VEC{r}_2-\VEC{x}_{23}\right) \nonumber \\
	\hspace{-0.25cm}&\times&\hspace{-0.25cm} 
	y_{L}^{M*}(\hat{x}_3)\delta n(\VEC{x}_1)\delta n(\VEC{x}_2)\delta n(\VEC{x}_3),
	\label{Eq:three_point_es}
\end{eqnarray}
where $\VEC{x}_{13}=\VEC{x}_1-\VEC{x}_3$, and $\VEC{x}_{23}=\VEC{x}_2-\VEC{x}_3$. The above expression can be rewritten into a FFT-based form
\begin{eqnarray}
	\widehat{\zeta}_{\ell_1\ell_2L}(r_1,r_2)
	\hspace{-0.25cm}&=&\hspace{-0.25cm} 
	 H_{\ell_1\ell_2L}\, \sum_{m_1m_2M}	  \left( \begin{smallmatrix} \ell_1 & \ell_2 & L \\ m_1 & m_2 & M \end{smallmatrix}  \right)  \nonumber \\
	\hspace{-0.25cm} &\times&\hspace{-0.25cm}
	\frac{N_{\ell_1\ell_2L}}{I}\int d^3x\,F_{\ell_1}^{\, m_1}(\VEC{x};r_1)\, F_{\ell_2}^{\, m_2}(\VEC{x};r_2)\,G_{L}^{M}(\VEC{x}), \nonumber \\
	\label{Eq:reduced_xi_estimator}
\end{eqnarray}
where
\begin{eqnarray}
	F_{\ell}^{\, m}(\VEC{x};r) \hspace{-0.25cm}&=& \hspace{-0.25cm}
	i^{\ell}\int \frac{dkk^2}{2\pi^2}j_{\ell}(rk) F_{\ell}^{\,m}(\VEC{x};k) \nonumber \\
	\hspace{-0.25cm}&=& \hspace{-0.25cm}
	i^{\ell}\int \frac{d^3k}{(2\pi)^3} 
	\mathrm{e}^{\mathrm{i}\VEC{k}\cdot\VEC{x}} j_{\ell}(rk) y_{\ell}^{m*}(\hat{k}) \frac{\delta n|_{\rm FFT}(\VEC{k})}{W_{\rm mass}(\VEC{k})}.
\end{eqnarray}
The $L=0$ mode of equation~(\ref{Eq:reduced_xi_estimator}) reproduces the estimator of the Legendre coefficients:
\begin{eqnarray}
	\widehat{\zeta}_{\ell}(r_1,r_2) = \widehat{\zeta}_{\ell\ell L=0}(r_1,r_2).
\end{eqnarray}
This $L=0$ mode estimator is the same as the FFT-based one presented in~\citet{Slepian:2016qwa}.

When we compute the three-point function using FFTs, we need to subtract the shot-noise term from the measured three-point function, just like in the case of the bispectrum. While there are the four shot-noise terms in the bispectrum, three of them, $S|_{i\neq j=k}$, $S|_{i=k\neq j}$, and $S|_{i=j=k}$, contribute to the three-point function at $r_1=0$, $r_2=0$, and $r_1=0$ and $r_2=0$, respectively. Therefore, we need to consider only one term coming from $S|_{i=j\neq k}$:
\begin{eqnarray}
	S_{\ell_1\ell_2L}\big|_{i=j\neq k}(r_1,r_2) 
	\hspace{-0.25cm}&=&\hspace{-0.25cm}
	 H_{\ell_1\ell_2L}\, \sum_{m_1m_2M}	  \left( \begin{smallmatrix} \ell_1 & \ell_2 & L \\ m_1 & m_2 & M \end{smallmatrix}  \right)  \nonumber \\
	\hspace{-0.25cm} &\times&\hspace{-0.25cm}
	\frac{\delta^{(\rm K)}_{r_1r_2}}{N_{\rm mode}(r)} \frac{N_{\rm grid}}{V_{\rm FFT}} 
	(-1)^{\ell_1+\ell_2} \frac{N_{\ell_1\ell_2L}}{I} \nonumber \\
	\hspace{-0.25cm}&\times&\hspace{-0.25cm}
	\int \frac{d^2\hat{r}_1}{4\pi} 
	  y_{\ell_1}^{m_1*}(\hat{r}_1)  y_{\ell_2}^{m_2*}(\hat{r}_1)  \int \frac{d^3k_3}{(2\pi)^3}\mathrm{e}^{\mathrm{i}\VEC{k}_3\cdot\VEC{r}_1}
	\nonumber \\
	\hspace{-0.25cm}&\times&\hspace{-0.25cm}
	\left[  \delta n_L^M(\VEC{k}_3) N_0^{0*}(\VEC{k}_3) -\frac{\bar{S}_L^M C_{\rm shot}(\VEC{k}_3)}{W_{\rm mass}^2(\VEC{k}_3)} \right],
\end{eqnarray}
where $V_{\rm FFT}$ is the volume of the Cartesian box in which the galaxies are placed before the FFT is performed, and $N_{\rm mode}(r)$ is the number of $r$-modes in each $r$-bin. Thus, the shot noise of the FFT-based estimator of the three-point function multipoles only contributes to the $r_1=r_2$ case.

\subsection{Estimators under the global plane-parallel approximation}
\label{Sec:GlobalLos}

In a situation like a cubic box simulation with periodic boundaries, we can define the global LOS direction and can choose the direction as the $z$-axis without loss of generality: the spherical harmonics with the LOS direction become $y_{L}^{M*}(\hat{n})=y_{L}^{M*}(\hat{z})=\delta_{M0}^{\rm (K)}$. Under this situation, we no longer need to compute the $y_L^{M*}$-weighted density fluctuations but the normal density fluctuation $\delta n=\delta n_{L=0}^{M=0}$, and the estimators of the bispectrum and three-point function multipoles are simplified to
\begin{eqnarray}
	\widehat{B}_{\ell_1\ell_2L}(k_1,k_2)
	\hspace{-0.25cm}&=& \hspace{-0.25cm}
	H_{\ell_1\ell_2L} \sum_{m}	  \left( \begin{smallmatrix} \ell_1 & \ell_2 & L \\ m & -m & 0 \end{smallmatrix}  \right) 
	 \nonumber \\
	\hspace{-0.25cm}  &\times&\hspace{-0.25cm}
	\frac{N_{\ell_1\ell_2L}}{I}\int d^3x\,   F_{\ell_1}^{\, m}(\VEC{x};k_1)\,F_{\ell_2}^{\, -m}(\VEC{x};k_2)\, G_{0}^0(\VEC{x}) \nonumber \\
	\widehat{\zeta}_{\ell_1\ell_2L}(r_1,r_2)
	\hspace{-0.25cm} &=& \hspace{-0.25cm}
	H_{\ell_1\ell_2L}
	  \sum_{m}	  \left( \begin{smallmatrix} \ell_1 & \ell_2 & L \\ m & -m & 0 \end{smallmatrix}  \right) 
	 \nonumber \\
	\hspace{-0.25cm}  &\times&\hspace{-0.25cm}
	\frac{N_{\ell_1\ell_2L}}{I}   \int d^3x\,   F_{\ell_1}^{\, m}(\VEC{x};r_1)\,
	   F_{\ell_2}^{\, -m}(\VEC{x};r_2)\,  G_{0}^0(\VEC{x}), \nonumber \\
	\label{Eq:simulation_estimator}
\end{eqnarray}
where the normalization factor $I$ is not represented by equation~(\ref{Eq:normalization}) but $I =(N_{\rm p}^3/V_{\rm box}^2)$ with $V_{\rm box}$ and $N_{\rm p}$ being the volume of a simulation box and the number of particles in the box, respectively.

Before closing this section, let us comment on an example of our computational time with our estimator.
We carried out numerical computations on XC50 at Center for Computational Astrophysics, National Astronomical Observatory of Japan, which consists of a suite of computational nodes with the CPU of Intel Xeon Gold 6148 (20 cores with 2.4-3.7GHz).

The FFT-based algorithm given by equation~(\ref{Eq:bi_estimator_FFTs}) requires some FFT operations for each combination of $k$-bins, $(k_1,k_2)$ and the summation of all possible $m_1$, $m_2$, and $M$ modes. Consequently, the computational complexity of our estimator is $\mathcal{O}\left(N_{\rm multipole} \times N_{\rm b}^2\times N_{\rm grid} \ln N_{\rm grid}\right)$, where $N_{\rm b}$ is the number of $k$-bins, and $N_{\rm multipole}$ depends on the bispectrum multipoles that we compute. 

For example, let us consider the lowest-order bispectrum multipole, $B_{000}(k_1,k_2)$, with $N_{\rm b}=11$ and $N_{\rm grid}=512^3$,
where the number of $k$-bin pairs of $B_{000}(k_1,k_2)$ is $N_{\rm b}(N_{\rm b}+1)/2 = 66$ because of the symmetry between $k_1$ and $k_2$.
We are then able to compute $B_{000}$ from a single mock (a single simulation box) in $1.22$ ($0.92$) CPU hours using our serial code.
As will be shown in Section~\ref{Sec:SignalNoiseRatio}, the diagonal part of $B_{000}$, i.e., $B_{000}(k_1=k_2,k_2)$, dominates the signal-to-noise of the bispectrum monopole measurements.
When we only compute this diagonal part that the number of $k$-bins is $11$,
measuring $B_{000}(k_1=k_2,k_2)$ takes $0.23$ ($0.2$) CPU hours for a mock (a box simulation), thus reducing enormously the computing time needed for the bispectrum measurements.
For the higher multipoles, their computational time can be estimated by multiplying that of $B_{000}$ by $N_{\rm multipole}$,
where $N_{\rm multipole}$ = $1$ ($1$), $3$ ($3$), $5$ ($5$), $5$ ($1$), $9$ ($3$), and $9$ ($1$) for $B_{000}$, $B_{110}$, $B_{220}$, $B_{202}$, $B_{112}$, and $B_{404}$ for the mock (the box simulation).

Now here is the estimate of computational time to complete the bispectrum measurements for the BOSS survey 
that spans in two distinct sky regions (North and South Galactic Caps) in three redshift bins ($0.3<z<0.5$, $0.4 < z < 0.6$, and $0.5<z<0.75$),
where each region has $2048$ MultiDark-Patchy mock catalogues~\cite{Kitaura:2015uqa,Klypin:2014kpa}.
Sticking to four multipoles, $B_{000}$, $B_{110}$, $B_{220}$, and $B_{202}$,
we can complete the bispectrum measurements in 
$(1+3+5+5) \times 6 \times 2048 \times 1.22 / 24 = 8745$ CPU days for the ``full'' bispectrum (i.e., $k_1\neq k_2$ is allowed) and $1650$ CPU days for the ``diagonal'' bispectrum (i.e., $(k_1=k_2)$ only).

\section{MEASUREMENTS}
\label{Sec:Measurements}

Having shown the new decomposition formalism of the bispectrum (Section~\ref{Sec:NewDecompositionFormalsim}) and its estimator (Section~\ref{Sec:Estimator}), we now turn to our primary goal to understand how the bispectrum multipoles provide cosmological information. 
In Section~\ref{Sec:Data}, we briefly introduce the galaxy data set, the random catalogues used for corrections of survey geometry effects, and the mock catalogues used to estimate the covariance matrix. 
After setting the properties of our 3D Cartesian grid in Section~\ref{Sec:Prescription}, we will measure the bispectrum multipoles of the CMASS NGC sample and the corresponding mock catalogues in Sections \ref{Sec:B000} and \ref{Sec:HigherMultipoles}. We then estimate the covariance matrix of the bispectrum multipoles in Section~\ref{Sec:CovarianceMatrix}. Finally, we compute cumulative signal-to-noise $(\rm {S/N})$ ratios estimated from the measurements of the bispectrum multipoles to see the impact of the bispectrum measurements on the information context of galaxy clustering in Section~\ref{Sec:SignalNoiseRatio}.

\subsection{DATA}
\label{Sec:Data}

We use the CMASS North Galactic Cap (NGC) sample with $586\,003$ galaxies in the redshift range 
$0.43 < z < 0.75$~\citep{White:2010ed,Parejko:2012ba,Bundy2015ApJS..221...15B,Leauthaud2016MNRAS.457.4021L,Saito2016MNRAS.460.1457S}. This sample is drawn from the Data Release 12~\citep[DR12;][]{Alam:2015mbd} of the Baryon Oscillation Spectroscopic Survey~\cite[BOSS;][]{Bolton:2012hz,Dawson:2012va}, which is part of the Sloan Digital Sky Survey III~\cite[SDSS-III;][]{Eisenstein:2011sa}, and is selected from multi-color SDSS imaging~\citep{Fukugita:1996qt,Gunn:1998vh,Smith:2002pca,Gunn:2006tw,Doi:2010rf}. We also use the associated random catalogues that quantify the survey geometry of BOSS.

To estimate the errors on the bispectrum, namely the covariance matrix, we utilize the MultiDark-Patchy mock catalogues~\cite[MD-Patchy mocks;][]{Kitaura:2015uqa,Klypin:2014kpa}, which are designed to reproduce the BOSS CMASS dataset. These mocks have been calibrated to an N-body based reference sample using approximate gravity solvers and analytical-statistical biasing models and incorporate observational effects including the survey geometry, veto mask, and fiber collisions. We decided to use all $2048$ mocks for the lowest order bispectrum multipole, $B_{000}$, and $600$ mocks for the other multipoles. The fiducial cosmology for these mocks assumes a $\Lambda$CDM cosmology with $(\Omega_{\rm \Lambda}, \Omega_{\rm m}, \Omega_{\rm b}, \sigma_8, h) = (0.692885, 0.307115, 0.048, 0.8288, 0.6777)$.

\subsection{Prescription for measurements}
\label{Sec:Prescription}

We use the TSC assignment function to bin the CMASS NGC dataset in a cubic box with a total volume of $(3.5\hGpc)^3$ split into $512^3$ grid cells. This corresponds to a grid-cell resolution of $\sim 6.8\hMpc$ and a Nyquist frequency of $k_{\rm N}=0.46\hk$. We then use the Fast Fourier Transform in the West (FFTW) library\footnote{ \url{http://fftw.org}} to perform the FFTs. Appendix~\ref{Ap:MassAssignmentFunction} provides a comparison of several assignment schemes, NGP, CIC, and TSC, with a different number of grid-cells ($512^3$ and $1024^3$), showing that the TSC scheme with $512^3$ grid cells achieves sub-percent accuracy for $k\lesssim 0.2\hk$.

To correct for several observational artifacts in the catalogues we use a completeness weight for each galaxy
~\citep{Ross:2012qm,Anderson:2013zyy,Reid:2015gra},
\begin{eqnarray}
	  w_{ {\rm c}}(\VEC{x}) = w_{ {\rm systot}}(\VEC{x}) \left( w_{ {\rm cp}}(\VEC{x}) + w_{ {\rm noz}}(\VEC{x}) - 1\right),
	  \label{Eq:weight_c}
\end{eqnarray}
where $\VEC{x}$ is the observed galaxy position, and $w_{\rm cp}$, $w_{\rm noz}$, and $w_{\rm systot}$ denote the redshift failure weight, the collision weight, and the angular systematics weight, respectively. The details about the observational systematic weights are described in~\citet{Reid:2015gra}. Additionally, we use a signal-to-noise weight, the so-called FKP weight, proposed by~\citealt{Feldman:1993ky}, $w_{\rm FKP}(\VEC{x})=1/\left[ 1 + \bar{n}(\VEC{x})P_0\right]$, with $P_0=10^4\, (\hMpc)^3$. For Gaussian errors, this weight function works even for higher order statistics~\citep{Scoccimarro:2000sn}. We investigate this weighting scheme in more detail in Appendix~\ref{Ap:FKPWeight}. By multiplying the completeness weight by the FKP weight, we finally define the local weight function that we use in our analysis: 
\begin{eqnarray}
	  w(\VEC{x}) = w_{\rm c}(\VEC{x})\, w_{\rm FKP}(\VEC{x}).
	  \label{Eq:weight}
\end{eqnarray}

We adopt a $k$-range of $0.01\hk\leq k\leq0.21\hk$ in bins of $\Delta k =0.02\hk$. The number of $k$-bins is 11 and the number of $k$-bin pairs for $B_{\ell_1\ell_2L}(k_1,k_2)$ is $11\times(11+1)/2=66$ if $\ell_1=\ell_2$, otherwise $11\times11=121$.

\subsection{Lowest order of the bispectrum multipoles}
\label{Sec:B000}

\begin{figure}
	\includegraphics[width=\columnwidth]{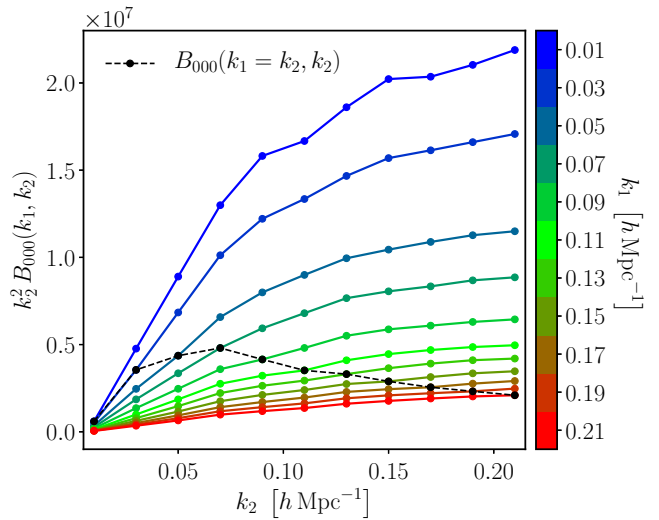}
	\caption{
    Means of the measurements of the bispectrum monopole, $B_{000}(k_1,k_2)$ (equation~\ref{Eq:bi_estimator_FFTs}), from the MD-Patchy mocks for CMASS NGC, with the shot noise terms (equation~\ref{Eq:ShotNoise}) subtracted.
	For plotting purposes, we fix $k_1$ to the value at each $k_1$-bin and plot $B_{000}(k_1,k_2)$ as a function of $k_2$.
	Since the number of $k_1$-bins is $11$, we plot $11$ colored lines:
	from top to bottom, they denote $B_{000}(k_1=0.01,k_2)$, $B_{000}(k_1=0.03,k_2)$, \dots, $B_{000}(k_1=0.21,k_2)$.
	We also plot the diagonal bispectrum monopole, $B_{000}(k_1=k_2,k_2)$ (black dashed line),  
	which intersects with $B_{000}(k_1,k_2)$ at each $k_2$-bin.
	}
	\label{fig:B000}
\end{figure}

We begin with measuring the lowest order of the bispectrum multipoles $B_{000}(k_1,k_2)$, which is hereafter referred to ``bispectrum monopole'', from the MD-Patchy mocks. In Fig.~\ref{fig:B000}, we show $B_{000}(k_1,k_2)$ measured using equation~(\ref{Eq:bi_estimator_FFTs}) after subtracting the shot noise terms (equation~\ref{Eq:ShotNoise}). For displaying purposes, we fix $k_1$ to the value at each $k_1$-bin: $k_1 = 0.01$, $0.03$, \dots, $0.21$, and plot $B_{000}(k_1,k_2)$ as a function of $k_2$. Each line in the figure is the mean bispectra in the mocks. From top (blue) to bottom (red) of the colored lines, the value of the fixed $k_1$ increases. For comparison, we also plot the mean of the diagonal elements of $B_{000}(k_1,k_2)$, i.e., $B_{000}(k_1=k_2,k_2)$ (black dashed), which intersects with $B_{000}(k_1,k_2)$ (colored lines) at each $k_2$-bin. In what follows, when we focus only on the diagonal elements, i.e., $B_{\ell_1\ell_2L}(k_1=k_2,k_2)$, we refer to them as ``diagonal bispectrum multipoles''. 

\subsection{Higher multipoles}
\label{Sec:HigherMultipoles}

\begin{figure*}
	\includegraphics[width=1.0\textwidth]{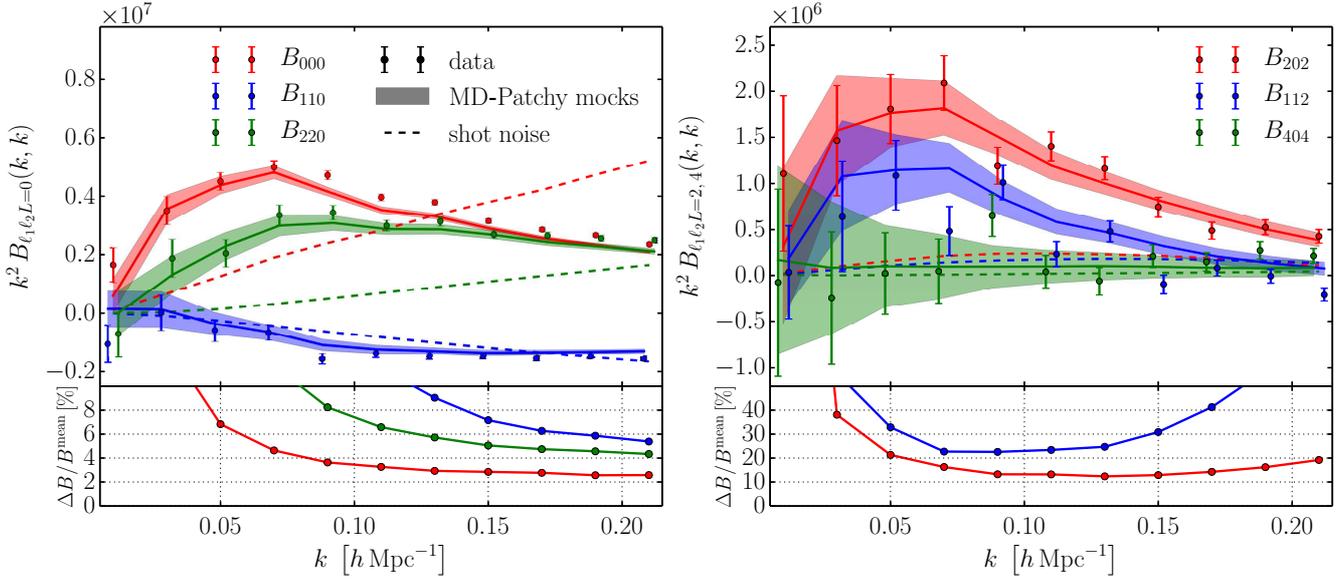}
	\caption{
	Diagonal bispectrum multipoles $B_{\ell_1\ell_2L}(k,k)$ 
	for the $L=0$ mode (left panel) and the $L\geq 2$ modes (right panel) measured from CMASS NGC,
    with the shot noise terms subtracted.
	The error bars on the data points are the $1\sigma$ errors, $\Delta B_{\ell_1\ell_2L}$, estimated from the MD-Patchy mocks.
	The shaded regions are the measurements from the MD-Patchy mocks with the $1\sigma$ errors. 
    For comparison, we plot the means of the shot noises measured from the MD-Patchy mocks for each bispectrum multipole (dashed lines).
	This figure demonstrates reasonable agreement between the MD-Patchy mocks and the observed galaxy sample at large scales $k \lesssim 0.1\hk$.
	The bottom parts in both the panels show the ratios of $\Delta B_{\ell_1\ell_2L}$ to the mean values of the measured $B_{\ell_1\ell_2L}$ in the mocks. 
	}
	\label{fig:bk}
\end{figure*}

We next turn to the higher order bispectrum multipoles, especially the $L\geq2$ modes that characterize the anisotropic signal, which is induced by the RSD or AP effects. Fig.~\ref{fig:bk} presents the measurements of the diagonal bispectrum multipoles from the CMASS NGC sample (data points). The error bars on the data points are the $1\sigma$ errors estimated from the MD-Patchy mocks (for details, see Section~\ref{Sec:CovarianceMatrix}). The solid lines denote the mean measured bispectrum multipoles in the MD-Patchy mocks, and the shaded regions are the $1\sigma$ errors on the mean values. As will be shown in Section~\ref{Sec:SignalNoiseRatio}, the diagonal elements of the bispectrum monopole mainly contribute to the ${\rm S/N}$ of the full bispectrum monopole. We expect similar results for higher bispectrum multipoles, and therefore, we focus on the diagonal bispectrum multipoles in this subsection.

The left panel in Fig.~\ref{fig:bk} displays the first three bispectrum multipoles of the $L=0$ mode: $B_{000}$, $B_{110}$, and $B_{220}$, where they correspond to the Legendre coefficients of the bispectrum, $B_{\ell=0,1,2}$ (equation~\ref{Eq:LegendreCoefficients}). While we find agreement between the results from the galaxy sample and the mocks at large scales ($k\lesssim0.1\hk$), they start to significantly depart from each other beyond the $1\sigma$ errors at small scales ($k\gtrsim0.1\hk$). We will ignore this tension in the subsequent analysis, and leave a more careful calibration of the mock catalogs to the measure bispectra for future work. 

In the right panel of Fig.~\ref{fig:bk}, we plot three bispectrum multiples of the $L\geq2$ modes: two in the $L=2$ mode; and one in the $L=4$ mode. As mentioned in Section~\ref{Sec:NewDecompositionFormalsim}, the $L=2$ and $4$ modes can be interpreted as the analogs of the power spectrum quadrupole and hexadecapole, respectively, because they mainly yield from the terms proportional to $(\hat{n}\cdot\VEC{v})^{L/2}$. As will be discussed in Section~\ref{Sec:SignalNoiseRatio} in a quantitative context, we can measure and detect the anisotropic signal in the bispectrum $L=2$ multipole with a high significance. On the other hand, the $L=4$ mode seems consistent with the null hypothesis of no signal. These results are similar to those in the power spectrum case in the BOSS analysis~\citep[e.g., see][]{Beutler:2016arn}:
while the power spectrum quadrupole is measured with high statistical significance, the power spectrum hexadecapole is not detected.

Fig.~\ref{fig:bk} also includes the means of the shot noise terms measured from the MD-Patchy mocks for each bispectrum multipole (dashed lines). For the $L=0$ mode, the shot noise terms are comparable to or larger than
the corresponding bispectrum multipoles at small scales ($k\sim0.2\hk$). This fact indicates that 
the shot noise dominates the errors on the L=0 mode measurements 
at $k\sim 0.2\hk$ in the BOSS analysis. Even for the $L>0$ modes,
the shot noise terms are not zero and significantly impact our measurements for all multipoles because of the anisotropic signals in the shot noise, i.e., $S_{\ell_1\ell_2L}|_{i\neq j = k}$, $S_{\ell_1\ell_2L}|_{i=k\neq j}$, and $S_{\ell_1\ell_2L}|_{i=j\neq k}$ in equation~(\ref{Eq:ShotNoise}).

The bottom parts in both panels of Fig.~\ref{fig:bk} show the ratios of the $1\sigma$ errors on the measured diagonal bispectrum multipole, $\Delta B_{\ell_1\ell_2L}$, to the means of $B_{\ell_1\ell_2L}$ measured in the MD-Patchy mocks. The values of these ratios depend on the bin width $\Delta k$, and smaller bin widths, e.g., $\Delta k = 0.01\hk$, will induce higher values because the number of $k$-modes at each bin decreases. Nevertheless, these plots provide a rough estimate of accuracy required in the modeling of the bispectrum multipoles. To use smaller scales than $k=0.1\hk$ for the bispectrum analysis, we should achieve $\sim 3\%$ accuracy for the theoretical model of $B_{000}$ and $\sim 10\%$ accuracy for $B_{202}$ in the BOSS analysis. 

\subsection{Covariance matrices}
\label{Sec:CovarianceMatrix}

\begin{figure*}
	\includegraphics[width=1.0\textwidth]{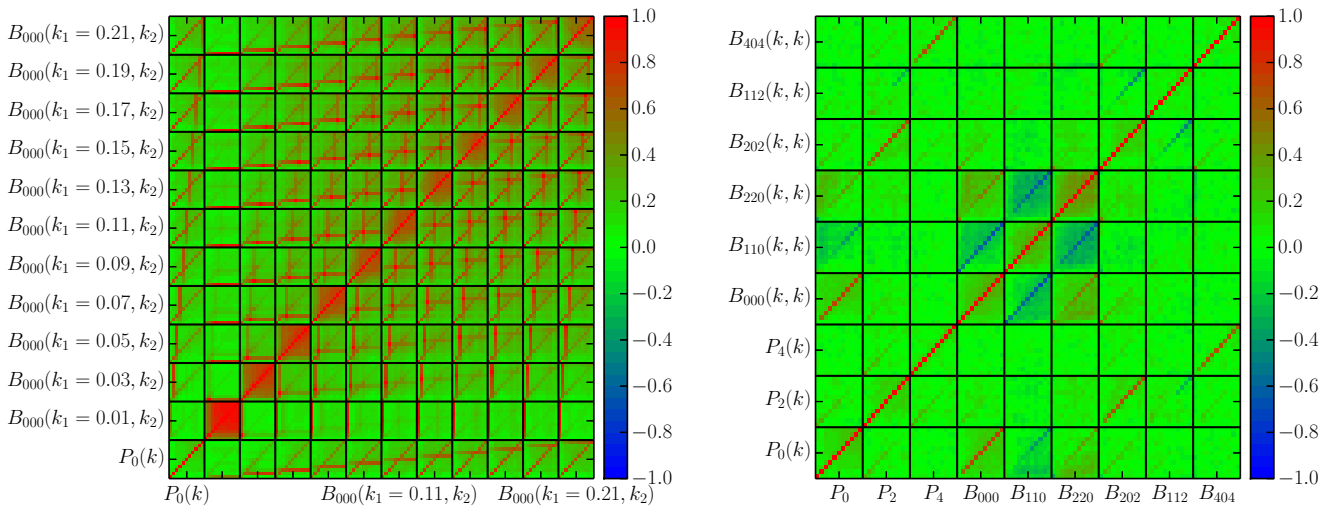}
	\caption{
Correlation matrices of the power spectrum and bispectrum multipoles, estimated from the MD-Patchy mocks for CMASS NGC. 
The left panel focuses on the power spectrum and bispectrum monopoles, $P_0(k)$ and $B_{000}(k_1,k_2)$;
the right shows the correlations among higher multipole terms of the power spectrum and the diagonal bispectrum.
Each block of both the panels separated by horizontal and vertical division lines contains $11$ $k$-bins in the $k$-range $k=0.01\, \mathchar`-\, 0.21\hk$.
The color indicates the level of correlation, where red and green represents high and low correlation, respectively. 
Remarkably, the diagonal bispectrum multipoles $B_{\ell_1\ell_2L}(k,k)$ estimated at different scales, i.e. at $k$ and $k'$ for $k\neq k'$, are nearly uncorrelated like the power spectrum case,
implying that the diagonal ones would provide dominant information on bispectrum measurements.
}
	\label{fig:cov_bk}
\end{figure*}

Now we move onto investigating the properties of the covariance matrix between the bispectrum multipoles, which describe statistical uncertainties of the bispectrum measurements. Constraining cosmological parameters, we commonly analyze the bispectrum in combination with the power spectrum, and hence, we also study the cross-covariance matrix between the power spectrum multipoles and the bispectrum multipoles to see how the bispectrum is correlated with the power spectrum.

Let $\VEC{X}$ be a data vector of measured quantities. The covariance matrix for $N_{\rm mock}$ mocks is then estimated by
\begin{eqnarray}
	\textbf{\textsf{C}} = \frac{1}{N_{\rm mock}-1} \sum_{n}^{N_{\rm mock}}
	\left( \VEC{X}^{(n)}-\bar{\VEC{X}} \right)^{\rm T}\left( \VEC{X}^{(n)}-\bar{\VEC{X}} \right),
	\label{Eq:Covariance}
\end{eqnarray}
where $\VEC{X}^{(n)}$ is the data vector obtained from the $n$-th mock, and the mean expectation value over the mocks is given by $\bar{\VEC{X}} = (1/N_{\rm mock})\sum_n^{N_{\rm mock}}\VEC{X}^{(n)}$. The correlation matrix $\textsf{\textbf{r}}$ is defined as $r_{ij} = C_{ij}/\left( C_{ii}C_{jj} \right)^{1/2}$, where $r_{ij}$ and $C_{ij}$ are the $ij$-element of the correlation matrix $\textsf{\textbf{r}}$ and the covariance matrix $\textsf{\textbf{C}}$, respectively, and the indexes $i$ and $j$ run over the number of bins. The case $i=j$ of the covariance, $C_{ii}$, gives the variance. The error bars shown in Fig.~\ref{fig:bk} are the square root of the variance, $C^{1/2}_{ii}$, of the data vector that consists of the diagonal bispectrum multipoles, $X_i=B_{\ell_1\ell_2L}(k_i,k_i)$.

First we focus on the power spectrum and bispectrum monopoles, namely $P_0$ and $B_{000}$. To represent $B_{000}(k_1,k_2)$ depending on $k_1$ and $k_2$, we fix $k_1$ to each $k_1$-bin value and describe it as a function of $k_2$ (see Section~\ref{Sec:B000}). We then consider the data set that consists of
\begin{eqnarray}
	\VEC{X}_1 \hspace{-0.25cm} &=&  \hspace{-0.25cm}  \{P_0(k), B_{000}(k_1=0.01,k_2), B_{000}(k_1=0.03,k_2), \nonumber \\
	      \hspace{-0.25cm}    &&  \hspace{-0.25cm} \dots, B_{000}(k_1=0.19,k_2), B_{000}(k_1=0.21,k_2)\},
	      \label{Eq:X1}
\end{eqnarray}
where each component of the data set, $P_0(k)$ or $B_{000}(k_1,k_2)$, contains $11$ $k$-bins in the $k$-range from $k=0.01\hk$ to $k=0.21\hk$. The left panel of Fig.~\ref{fig:cov_bk} displays the correlation matrix of the above data set, which is divided into $12\times12$ blocks by $11$ horizontal and vertical division lines. Each block contains $11\times11$ elements because of the $11$ $k$-bins. The covariance matrix of the bispectrum monopole depends on four variables: $k_1$, $k_2$, $k'_1$, and $k'_2$. We find from the left panel that there are strong correlations among the bispectrum monopoles estimated at given scales if $k_1=k'_2$ or $k_1=k'_2$, i.e., between $B_{000}(k_1,k_2)$ and $B_{000}(k_1,k'_2)$ and between $B_{000}(k_1,k_2)$ and $B_{000}(k'_1,k_2)$. Similarly, the bispectrum monopole $B_{000}(k_1,k_2)$ and the power spectrum monopole $P_0(k)$ are correlated if $k_1=k$ or $k_2=k$, i.e., between $B_{000}(k_1,k_2)$ and $P(k_1)$ and between $B_{000}(k_1,k_2)$ and $P(k_2)$. Strong correlation is observed whenever one of the two bispectrum scales $k_1$ and $k_2$ is matched with one of the scales of the other data vector. These results demonstrate that using only a few terms of $B_{000}(k_1,k_2)$ with different values of $k_1$ is a poor choice, and the full correlations considering all $k_1$ values should be used for any quantitative analysis.

Secondly, we consider the correlation matrix including higher multipoles of the power spectrum and bispectrum. Here, we only use the diagonal bispectrum multipoles $B_{\ell_1\ell_2L}(k,k)$. The data set then consists of
\begin{eqnarray}
	\VEC{X}_2  \hspace{-0.25cm} &=&  \hspace{-0.25cm}  \{P_0(k), P_2(k), P_4(k), B_{000}(k,k), B_{110}(k,k), \nonumber \\
	          \hspace{-0.25cm}  &&  \hspace{-0.25cm} B_{220}(k,k), B_{202}(k,k), B_{112}(k,k), B_{404}(k,k)\}.
\end{eqnarray}
The right panel of Fig.~\ref{fig:cov_bk} shows the correlation matrix of $\VEC{X}_2$, which is divided into $9\times9$ blocks. A remarkable feature in the panel is that the values of off-diagonal elements in each block are very small, implying that the estimates of the diagonal bispectrum multipoles are nearly uncorrelated unless $k=k'$. We can reproduce this characteristic feature using  analytical calculations in perturbation theory (Sugiyama et al. in preparation). We expect from this result that the diagonal bispectrum multipoles dominate the signal-to-noise ratio on bispectrum measurements (see Section~\ref{Sec:SignalNoiseRatio}). As expected in Section~\ref{Sec:NewDecompositionFormalsim}, we find strong correlations between $P_{\ell}$ and $B_{\ell_1\ell_2L}$ if $\ell=L$, otherwise weak correlations.

\subsection{Signal-to-noise ratios}
\label{Sec:SignalNoiseRatio}

\begin{figure}
	\includegraphics[width=\columnwidth]{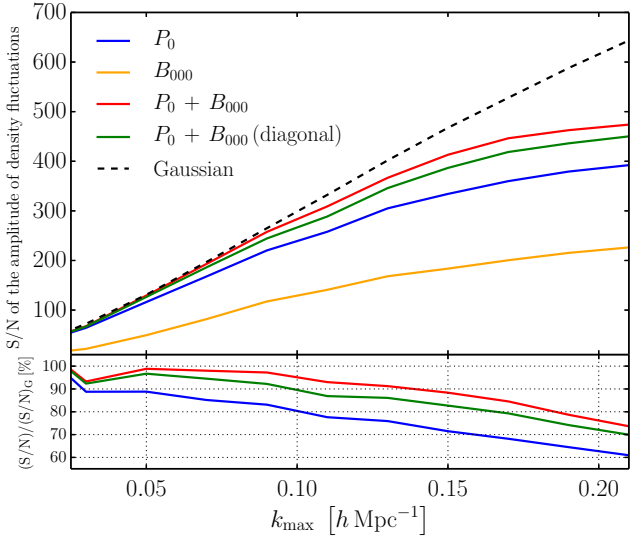}
	\caption{Cumulative signal-to-noise ratios of the amplitude of the measured density fluctuations from CMASS NGC as a function of $k_{\rm max}$. The blue line shows the S/N derived from the power spectrum monopole alone. The orange line shows the bispectrum monopole alone. Combining the power spectrum and bispectrum monopoles leads to the red line.
	The black dashed line shows the theoretical prediction using the Gaussian approximation, i.e., the linear power spectrum alone.
	For comparison, we also plot the result from the joint analysis using the diagonal bispectrum monopole, $B_{000}(k,k)$ (green solid line).
	The bottom part shows the ratios of the estimated ${\rm S/Ns}$ to the Gaussian prediction.
	This figure shows that
	while the ${\rm S/N}$ of the power spectrum alone is degraded to $\sim80\%$ at $k_{\rm max}=0.1\hk$ and $\sim60\%$ at $k_{\rm max}=0.2\hk$
	compared to the Gaussian prediction, the additional information of the bispectrum recovers the S/N up to $\sim95\%$ at $k_{\rm max}=0.1\hk$ and $\sim75\%$ at $k_{\rm max}=0.2\hk$.
	Furthermore, we find that the joint analysis using only the diagonal elements of the bispectrum monopole (green line) already gets very close to the red line,
	indicating that the diagonal bispectrum provides dominant information on the bispectrum measurements.
	}
	\label{fig:TotalSN}
\end{figure} %

\begin{figure*}
	\includegraphics[width=1.0\textwidth]{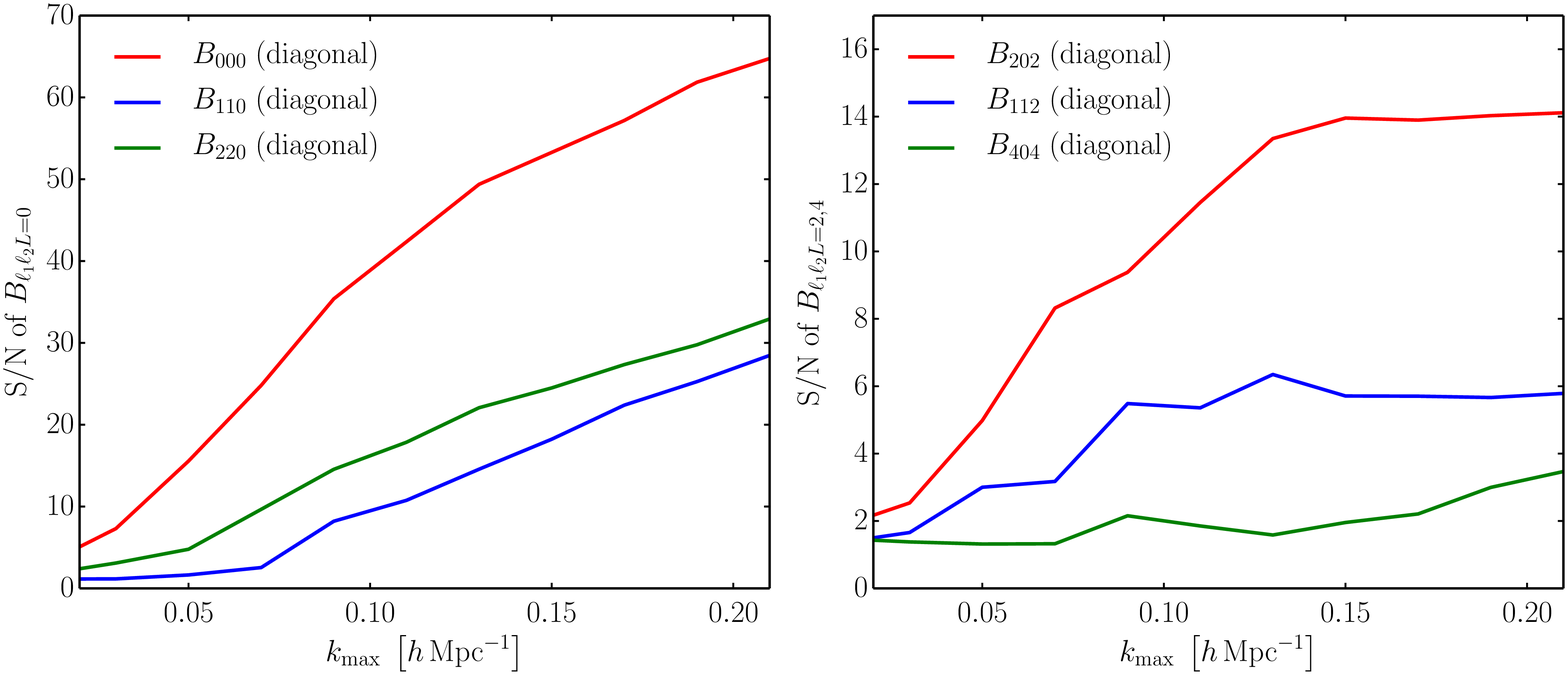}
	\caption{
		Cumulative ${\rm S/Ns}$ of the diagonal bispectrum multipoles.
		The left panel shows the ${\rm S/Ns}$ for the $L=0$ mode of the bispectrum multipoles;
		the right panel plots those for the $L\geq2$ modes.
		For each $L=0$ and $L=2$ mode, the lowest order of the bispectrum multipoles, $B_{000}$ ($L=0$) or $B_{202}$ ($L=2$),
		provides the highest ${\rm S/N}$ in the $L$ mode. This figure shows a $14\sigma$ detection of $B_{202}$, which is induced only by anisotropic signals on the bispectrum (the RSD or AP effects). 
	}
	\label{fig:SN}
\end{figure*}

A useful way to quantify the information context of galaxy clustering is to estimate cumulative signal-to-noise (S/N) ratios associated with the galaxy clustering measurements. The feature of the cumulative S/N has been well studied using simulations for the power spectrum alone~\citep{Rimes:2005xs,Rimes:2006dz,Takahashi:2009bq,Blot:2014pga}, and for a combination of the power spectrum and the bispectrum~\citep{Chan:2016ehg,Byun:2017fkz}. In this subsection, we compute the ``observed'' ${\rm S/Ns}$ by fitting the measured bispectrum multipoles from CMASS NGC to those from the MD-Patchy mocks.

\subsubsection{Joint analysis of the power spectrum and bispectrum monopoles}
\label{Sec:Monopole}

To illustrate how the bispectrum measurements provide cosmological information on galaxy clustering in addition to the power spectrum measurements, we estimate the cumulative S/N of the amplitude of density fluctuations in the joint analysis of the power spectrum and bispectrum monopoles. Namely, we assume that the observed density contrast $\delta_{\rm obs}$ is expressed by its theoretical prediction $\delta_{\rm theory}$ times a free parameter $\alpha$: $\delta_{\rm obs}=\alpha\,\delta_{\rm theory}$, and define the ${\rm S/N}$ as the statistical significance of $\alpha$. We then fit the measurements of $P_0$ and $B_{000}$ from CMASS NGC to $\alpha^2 P^{\rm mean}_{0}$ and $\alpha^3 B^{\rm mean}_{000}$, where $P^{\rm mean}_{\ell}$ and $B^{\rm mean}_{\ell_1\ell_2L}$ are the mean values in the MD-Patchy mocks. We compute the covariance matrices of $P_0$ and $B_{000}$ from the MD-Patchy mocks with the fiducial value of $\alpha$ being unity. We perform the standard likelihood analysis through the likelihood function $L\propto \exp\left( -\chi^2/2 \right)$ with the $\chi^2$ statistics and estimate the mean of $\alpha$, which is represented as $\langle\alpha\rangle$, and its $1\sigma$ error, $\Delta \alpha$, with the flat prior $0\leq \alpha \leq 10$. In this analysis, we account for a Hartlap factor~\citep{Hartlap:2006kj} and a Percival factor~\citep{Percival:2013sga} (for details, see Appendix~\ref{Ap:HartlapAndPercival}). Finally, we define the S/N as ${\rm S/N}\equiv \langle\alpha\rangle/\Delta \alpha$. We show the ${\rm S/N}$ as a function of the maximum wavenumber $k_{\rm max}$ with fixing the minimum wavenumber to $k_{\rm min}=0.01\hk$. For comparison, we also conduct the same analysis using only $P_0$ or $B_{000}$.

Fig.~\ref{fig:TotalSN} compares the ${\rm S/Ns}$ computed by several analyses. The blue solid line shows the analysis considering only the power spectrum monopole, the orange solid line represents the analysis using only the bispectrum monopole, and the red solid line denotes the joint analysis of the power spectrum and bispectrum monopoles, where the data set used in the joint analysis is given by equation~(\ref{Eq:X1}). For comparison, we also plot a theoretical prediction using the power spectrum monopole including the shot-noise term in the Gaussian limit (black dashed). The bottom part of this figure shows the ratios of the estimated ${\rm S/Ns}$ to the Gaussian prediction. If the galaxy distribution is entirely Gaussian, the power spectrum encodes a complete description of galaxy clustering. In this sense, the black dashed line shows the upper limit of the ${\rm S/N}$ that we can obtain from the CMASS NGC sample. In fact, we observe the well known non-linear degradation of the ${\rm S/N}$ of the power spectrum.  The degradation is $\sim 80\%$ at $k_{\rm max}=0.1\hk$ and $\sim60\%$ at $k_{\rm max}=0.2\hk$ compared to the Gaussian prediction. We find that the additional information of the bispectrum monopole recovers the ${\rm S/N}$ up to $\sim95\%$ at $k_{\rm max}=0.1\hk$ and $\sim75\%$ at $k_{\rm max}=0.2\hk$.

We also consider another joint analysis using the diagonal bispectrum monopole (the green solid line shown in Fig.~\ref{fig:TotalSN}), where the data set we use consists of $\VEC{X}=\{P_0(k), B_{000}(k,k)\}$. We stress that the number of bins of the diagonal bispectrum monopole, which is $11$, is smaller than that of the full bispectrum monopole depending on $k_1$ and $k_2$ by a factor of $6$. Nevertheless, this analysis well recovers the ${\rm S/N}$ up to $\sim90\%$ at $k_{\rm max}=0.1\hk$ and $\sim70\%$ at $k_{\rm max}=0.2\hk$. This is because the diagonal bispectrum allows us to extract nearly independent information at different scales (Section~\ref{Sec:CovarianceMatrix}). We conclude from this result that the diagonal bispectrum monopole dominates the signal-to-noise of the bispectrum monopole measurements, and we expect similar results for higher bispectrum multipoles.

\subsubsection{Bispectrum multipoles}

Also of interest is the detectability of the bispectrum multipoles themselves, especially for the $L\neq 0$ modes arising from the RSD or AP effects. To show this, we apply the same analysis as done in Section~\ref{Sec:Monopole} to the diagonal bispectrum multipoles, where we adopt the theoretical templates as $\beta B_{\ell_1\ell_2L}^{\rm mean}$ with $\beta$ being a varying parameter, and the ${\rm S/N}$ of the bispectrum multipoles is then defined as ${\rm S/N} = \langle \beta \rangle/\Delta \beta$.

The left panel of Fig.~\ref{fig:SN} shows the ${\rm S/Ns}$ estimated from the first three multipoles of the $L=0$ mode: $B_{000}$ (red), $B_{110}$ (blue), and $B_{220}$ (green), whose measurements are displayed by the same colored data points with the $1\sigma$ error bars in the left panel of Fig.~\ref{fig:bk}. As expected, the largest ${\rm S/N}$ arises from the bispectrum monopole. The ${\rm S/Ns}$ of both $B_{110}$ and $B_{220}$ are about half of the ${\rm S/N}$ of $B_{000}$. Interestingly, we find the ${\rm S/N}$ of $B_{110}$ to be smaller than that of $B_{220}$. This is because the amplitude of $B_{110}$ is about $1.5$ times smaller than $B_{220}$ at e.g. $k=0.2\hk$ (Fig.~\ref{fig:bk}), while the $1\sigma$ error of $B_{110}$ is about $1.25$ times smaller than $B_{220}$ at that scale. 

The right panel of Fig.~\ref{fig:SN} shows the ${\rm S/Ns}$ of three diagonal bispectrum multipoles with $L\geq2$: $B_{202}$ (red), $B_{112}$ (blue), and $B_{404}$ (green), whose measurements are displayed in the right panel of Fig.~\ref{fig:bk}. We find that the ${\rm S/Ns}$ of these $L>0$ modes are all smaller than those of the first three multipoles of the $L=0$ mode. For the $L=2$ mode, the ${\rm S/N}$ of $B_{202}$ is about two times as large as that of $B_{112}$, showing that $B_{202}$ dominates the signal-to-noise of the $L=2$ mode measurements. The lowest order of the $L=4$ mode, $B_{404}$, is consistent with the null hypothesis of no signal: ${\rm S/N} < 2$ up to $k_{\rm max}=0.15\hk$ and ${\rm S/N}<3.5$ at $k_{\rm max}=0.21\hk$.

The most striking feature of the right panel is that the ${\rm S/N}$ of $B_{202}$ is $14$ at $k_{\rm max}>0.15\hk$, 
corresponding to a $14\sigma$ detection of the anisotropic bispectrum signal in the lowest order $L=2$ mode. We have also detected the next leading order $L=2$ mode, $B_{112}$, at the $6\sigma$ level. Since we use the mean bispectra measured from the MD-Patchy mocks as the theoretical template to fit the measurements, we ignore the AP effect in this analysis. Nevertheless, we can well fit the measurements with
$\chi^2/{\rm d.o.f}=9.3/7$ for $B_{202}$ at $k_{\rm max}=0.15\hk$ and $\chi^2/{\rm d.o.f}=11.5/6$ for $B_{112}$ at $k_{\rm max}=0.13\hk$, indicating that the anisotropic signals detected here are well explained by our models of the RSD effect alone without the AP effect, i.e., by the mean bispectra measured from the MD-Patchy mocks.

\section{SURVEY WINDOW FUNCTIONS}
\label{Sec:SurveyWindowFunctions}

\begin{figure*}
	\includegraphics[width=1.0\textwidth]{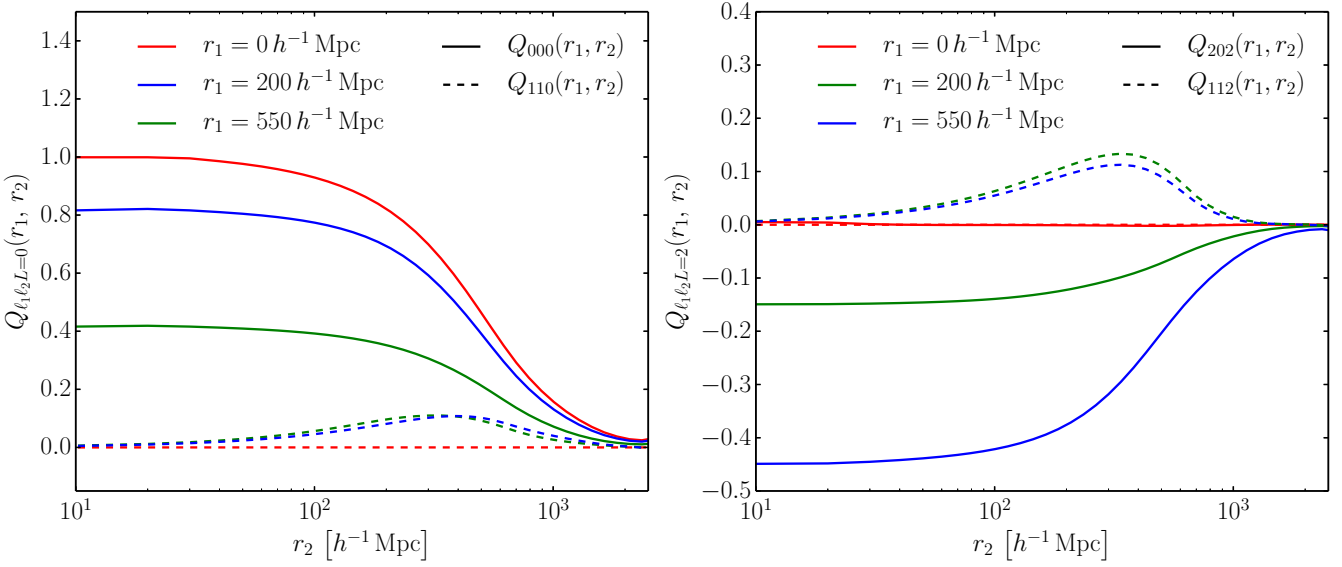}
	\caption{
Multipole components of the three-point survey window function, $Q_{\ell_1\ell_2L}(r_1,r_2)$, as given in equation~(\ref{Eq:Q}), which are estimated for survey geometry of CMASS NGC.
For display purposes, we fix $r_1$ to $0$ (red), $200$ (green), and $550\hMpc$ (blue), and plot $Q_{\ell_1\ell_2L}(r_1,r_2)$ as a function $r_2$.
We show four components: $Q_{000}$ (solid) and $Q_{110}$ (dashed) in the left panel; $Q_{202}$ (solid) and $Q_{112}$ (dashed) in the right panel.
At large scales, all of the window function multipoles becomes zero, while at small scales, they become constant if $\ell_2=0$, otherwise zero.
	}
	\label{fig:window}
\end{figure*}

In this section, we discuss the effects of survey geometry on the bispectrum. There has been a lot of effort to include the survey geometry effect in Fourier-space data analysis because the survey geometry distorts the observed density fluctuation and can introduce spurious anisotropic signals. To quantify the survey geometry distortion, \citet{Beutler:2013yhm}, for the first time, proposed to expand the survey window functions in Legendre polynomials and defined the window function multipoles. This treatment was further developed by~\citet{Wilson:2015lup}, based on configuration-space calculations. \citet{Sugiyama:2017ggb} extended the \citet{Wilson:2015lup} formalism to the Bipolar spherical harmonic expansion formalism to explore the rotational invariance breaking of the survey geometry. We apply these formalisms, that have been developed in the power spectrum analysis, to the bispectrum analysis and present a framework how one can correct for the survey geometry effect when theoretically modeling the bispectrum. 

The outline of this section is as follows. In Section~\ref{Sec:WindowFunctionMultipoles}, we define a three-point window function in configuration space and decompose it into the TripoSH basis with zero total angular momentum like the three-point function multipoles given in equation~(\ref{Eq:3pcfmultipoles}). In Section~\ref{Sec:MaskedBispectrumMultipoles}, we show the derivation of the bispectrum multipoles including the survey window corrections. In Section~\ref{Sec:ComparisonWithSimulations}, we compute the masked bispectrum multipoles in the second-order perturbation theory and compare them with the measurements from simulations with and without the survey geometry effect.

\subsection{Window function multipoles}
\label{Sec:WindowFunctionMultipoles}

We define the multipole components of the three-point window function in configuration space by replacing the density fluctuation $\delta n$ (equation~\ref{Eq:three_point_es}) by the mean number density measured from a random catalogue $\bar{n}$ (equation~\ref{Eq:mean_n}):
\begin{eqnarray}
	Q_{\ell_1\ell_2L}(r_1,r_2)
	\hspace{-0.25cm}&=&\hspace{-0.25cm}
	 H_{\ell_1\ell_2L}\, \sum_{m_1m_2M}	  \left( \begin{smallmatrix} \ell_1 & \ell_2 & L \\ m_1 & m_2 & M \end{smallmatrix}  \right)  \nonumber \\
	\hspace{-0.25cm} &\times&\hspace{-0.25cm}
	\frac{N_{\ell_1\ell_2L}}{I}\int \frac{d^2\hat{r}_1}{4\pi}y_{\ell_1}^{m_1*}(\hat{r}_1)\int \frac{d^2\hat{r}_2}{4\pi}y_{\ell_2}^{m_2*}(\hat{r}_2)
	\nonumber \\
	\hspace{-0.25cm}&\times&\hspace{-0.25cm} 
	\int d^3x_1\int d^3x_2\int d^3x_3  \nonumber \\
	\hspace{-0.25cm}&\times&\hspace{-0.25cm} 
	\delta_{\rm D}\left( \VEC{r}_1-\VEC{x}_{13}\right) \delta_{\rm D}\left( \VEC{r}_2-\VEC{x}_{23}\right) \nonumber \\
	\hspace{-0.25cm}&\times&\hspace{-0.25cm} 
	y_{L}^{M*}(\hat{x}_3)\,\bar{n}(\VEC{x}_1)\,\bar{n}(\VEC{x}_2)\,\bar{n}(\VEC{x}_3).
	\label{Eq:Q}
\end{eqnarray}
These three-point window function multipoles can be computed by the same algorithm used to measure the three-point function multipoles, allowing us to use FFTs. 

In this work, we measure four components of $Q_{\ell_1\ell_2L}$ from the random catalogue associated with CMASS NGC and plot them in Fig.~\ref{fig:window}: $Q_{000}$ (solid) and $Q_{110}$ (dashed) of the $L=0$ mode in the left panel; $Q_{202}$ (solid) and $Q_{112}$ (dashed) of the $L=2$ mode in the right panel. To display the window function multipoles depending on two comoving distances $r_1$ and $r_2$, we fix $r_1$ to $r_1=0$ (red), $200$ (green), and $550\hMpc$ (blue), and plot $Q_{\ell_1\ell_2L}(r_1,r_2)$ as a function of $r_2$. On large scales going beyond the survey regime, all of the window function multipoles should become zero, because we cannot find pairs (triplets) of galaxies at these scales, which is the so-called survey edge effect. On the other hand, on small scales where the survey edge effect no longer matters, the window function multipoles that depend only on $r_2$ after fixing $r_1$ will be constant if $\ell_2=0$ or zero for $\ell_2\geq1$. In particular, $Q_{000}(r_1=0,\, r_2)$, by definition, approaches unity in the limit $r_2 \rightarrow 0$. Since the window function multipoles are symmetric under $r_1\leftrightarrow r_2$ and $\ell_1\leftrightarrow \ell_2$, i.e., $Q_{\ell_1\ell_2L}(r_1,r_2) = Q_{\ell_2\ell_1L}(r_2,r_1)$,  
for example, $Q_{000}(r_1=550,\, r_2)$ does not become unity at $r_2=0$ but becomes the same as $Q_{000}(r_1=0,\,r_2=550)$.
We expect similar results for all of the other multipoles and any survey regions such as CMASS SGC, LOWZ NGC, and LOWZ SGC, even though their exact shapes will depend on each survey geometry.

\subsection{Masked bispectrum multipoles}
\label{Sec:MaskedBispectrumMultipoles}

Now we are ready to derive the bispectrum multipoles including survey window corrections. To do so, we first compute the ensemble average of the estimator of the three-point function multipoles (equation~\ref{Eq:three_point_es}):
\begin{eqnarray}
	\left\langle \widehat{\zeta}_{\ell_1\ell_2L}(r_1,r_2)\right\rangle
	\hspace{-0.25cm} &=& \hspace{-0.25cm}  
	H_{\ell_1\ell_2L} \sum_{m_1m_2M}  \left( \begin{smallmatrix} \ell_1 & \ell_2 & L \\ m_1 & m_2 & M \end{smallmatrix}  \right)  \nonumber \\
	\hspace{-0.25cm}&\times&\hspace{-0.25cm} 
	\frac{N_{\ell_1\ell_2L}}{I},\int \frac{d^2\hat{r}_1}{4\pi}y_{\ell_1}^{m_1*}(\hat{r}_1)\int \frac{d^2\hat{r}_2}{4\pi}y_{\ell_2}^{m_2*}(\hat{r}_2) \nonumber \\
	\hspace{-0.25cm}&\times&\hspace{-0.25cm} 
	\int d^3x_1\int d^3x_2\int d^3x_3\, y_{L}^{M*}(\hat{x}_3) \nonumber \\
	\hspace{-0.25cm}&\times&\hspace{-0.25cm} 
	\delta_{\rm D}\left( \VEC{r}_1-\VEC{x}_{13}\right) \delta_{\rm D}\left( \VEC{r}_2-\VEC{x}_{23}\right) \nonumber \\
	\hspace{-0.25cm}&\times&\hspace{-0.25cm} 
	\left\langle \delta n(\VEC{x}_1)\,\delta n(\VEC{x}_2)\,\delta n(\VEC{x}_3) \right\rangle .
	\label{Eq:Ensemble}
\end{eqnarray}
The theoretical expression of the three-point function of the observed density fluctuation $\delta n$ (equation~\ref{Eq:dn}) is described as
\begin{eqnarray}
	  &&\left\langle \delta n(\VEC{x}_1)\,\delta n(\VEC{x}_2)\,\delta n(\VEC{x}_3) \right\rangle \nonumber \\
	  &=& \bar{n}(\VEC{x}_1) \bar{n}(\VEC{x}_2) \bar{n}(\VEC{x}_3)\left[\, \zeta(\VEC{x}_{13},\VEC{x}_{23}, \hat{x}_{3})- \bar{\zeta}\,  \right],
	  \label{Eq:zeta_obs_theory}
\end{eqnarray}
where we applied the local plane-parallel approximation and chose $\hat{x}_3$ as the LOS direction. $\bar{\zeta}$, which is the so-called integral constraint~\citep{Peacock1991}, which comes from the difference between the measured mean density from a finite survey volume and the true value. We determine the integral constraint to satisfy the condition:
\begin{eqnarray}
	\int dr_1 r_1^2\int dr_2 r_2^2\, \left\langle \widehat{\zeta}_{000}(r_1,r_2)\right\rangle = 0.
\end{eqnarray}
The three point function depending on two relative coordinates, $\VEC{x}_{13}$ and $\VEC{x}_{23}$, and one LOS direction, $\hat{x}_3$, can be decomposed into the TripoSH basis with zero total angular momentum
(equation~\ref{Eq:Zeta_TripoSH_J0}):
\begin{eqnarray}
	\hspace{-0.8cm}&&\zeta(\VEC{x}_{13},\VEC{x}_{23}, \hat{x}_{3}) \nonumber \\
	\hspace{-0.25cm}&=&\hspace{-0.45cm} \sum_{\ell_1+\ell_2+L={\rm even}}\hspace{-0.4cm} \zeta_{\ell_1\ell_2L}(|\VEC{x}_{13}|,|\VEC{x}_{23}|)\,S_{\ell_1\ell_2L}(\hat{x}_{13},\hat{x}_{23},\hat{x}_3).
	\label{Eq:3ptcTripoSH}
\end{eqnarray}
Inserting equations~(\ref{Eq:zeta_obs_theory}) and (\ref{Eq:3ptcTripoSH}) in equation~(\ref{Eq:Ensemble}), and using equations~(\ref{Eq:YY2Y}) and (\ref{Eq:Wigner9j}), we obtain the masked three-point function multipoles:
\begin{eqnarray}
	&&\left\langle \widehat{\zeta}_{\ell_1\ell_2L}(r_1,r_2)\right\rangle_{\rm model} \nonumber \\
	\hspace{-0.25cm}&=&\hspace{-0.25cm} 
	N_{\ell_1\ell_2L}\, \sum_{\ell'_1+\ell'_2+L'={\rm even}}\ \ \sum_{\ell''_1+\ell''_2+L''={\rm even} } \nonumber \\
	\hspace{-0.25cm}&\times&\hspace{-0.25cm} 
	\left\{ \begin{smallmatrix} \ell''_1 & \ell''_2 & L'' \\   \ell'_1 & \ell'_2 & L' \\   \ell_1 & \ell_2 & L\end{smallmatrix}  \right\}
	\left[\frac{H_{\ell_1\ell_2L}H_{\ell_1\ell'_1\ell''_1}H_{\ell_2\ell'_2\ell''_2}H_{L L'L''}  }{H_{\ell'_1\ell'_2L'}H_{\ell''_1\ell''_2L''}} \right]
	\nonumber \\
	\hspace{-0.25cm}&\times&\hspace{-0.25cm} 
			Q_{\ell''_1\ell''_2L''}(r_1,r_2)\, \zeta_{\ell'_1\ell'_2L'}(r_1,r_2) \nonumber \\
	&-& Q_{\ell_1\ell_2L}(r_1,r_2)\,\bar{\zeta},
	\label{Eq:zetaMask}
\end{eqnarray}
where "model" means that this masked model will be compared with the measured estimator.
The bracket with 9 multipole indices, $\{\dots\}$, denotes the Wigner-9j symbol whose definition is given in equation~(\ref{Eq:Wigner9j}).
The integral constraint in the above expression is given by
\begin{eqnarray}
	\bar{\zeta}
	\hspace{-0.25cm}&=&\hspace{-0.25cm}
	\frac{1}{\int dr_1 r_1^2\int dr_2 r_2^2\, Q_{000}(r_1,r_2)}\sum_{\ell_1+\ell_2+L={\rm even}}\nonumber \\
	\hspace{-0.25cm}&&\hspace{-0.25cm}
	\int dr_1 r_1^2\int dr_2 r_2^2  \frac{Q_{\ell_1\ell_2L}(r_1,r_2)\zeta_{\ell_1\ell_2L}(r_1,r_2)}{H_{\ell_1\ell_2L}^2N_{\ell_1\ell_2L}}.
	\label{Eq:IC}
\end{eqnarray}

We can schematically represent the masked three-point function multipoles as follows:
\begin{eqnarray}
	\left\langle \widehat{\zeta}_{\ell_1\ell_2L}(r_1,r_2)\right\rangle_{\rm model}
	\hspace{-0.25cm}&=&\hspace{-0.25cm}  Q_{000}(r_1,r_2)\, \zeta_{\ell_1\ell_2L}(r_1,r_2) - Q_{\ell_1\ell_2L}(r_1,r_2)\, \bar{\zeta}\nonumber \\
	\hspace{-0.25cm}&+&\hspace{-0.25cm} \left[  \mbox{the other multipole components}  \right].
	\label{Eq:zetalllMask}
\end{eqnarray}
On the right-hand side of the above expression, the first term, $Q_{000}\zeta_{\ell_1\ell_2L}$, is directly related to what we measure via the estimator $\widehat{\zeta}_{\ell_1\ell_2L}$. This term only includes the survey edge effect (Fig.~\ref{fig:window}). The second term, $Q_{\ell_1\ell_2L}\bar{\zeta}$, is the integral constraint correction. The survey window function does not only distort the individual multipoles, but also correlates different multipoles (the second line in the above expression). For an example, the masked three-point function monopole is given by (see also Appendix~\ref{Ap:WindowCorrections})
\begin{eqnarray}
   \left \langle \widehat{\zeta}_{000}(r_1,r_2) \right\rangle_{\rm model}\hspace{-0.25cm}&=&\hspace{-0.25cm} Q_{000}(r_1,r_2)\, \left[\,\zeta_{000}(r_1,r_2)-\bar{\zeta}\,\right] \nonumber \\
   \hspace{-0.25cm}&+&\hspace{-0.25cm} \frac{1}{3}\, Q_{110}(r_1,r_2)\, \zeta_{110}(r_1,r_2) \nonumber \\
   \hspace{-0.25cm}&+&\hspace{-0.25cm} \frac{1}{5}\, Q_{220}(r_1,r_2)\, \zeta_{220}(r_1,r_2) + \cdots.
\end{eqnarray}

\begin{figure*}
	\includegraphics[width=1.0\textwidth]{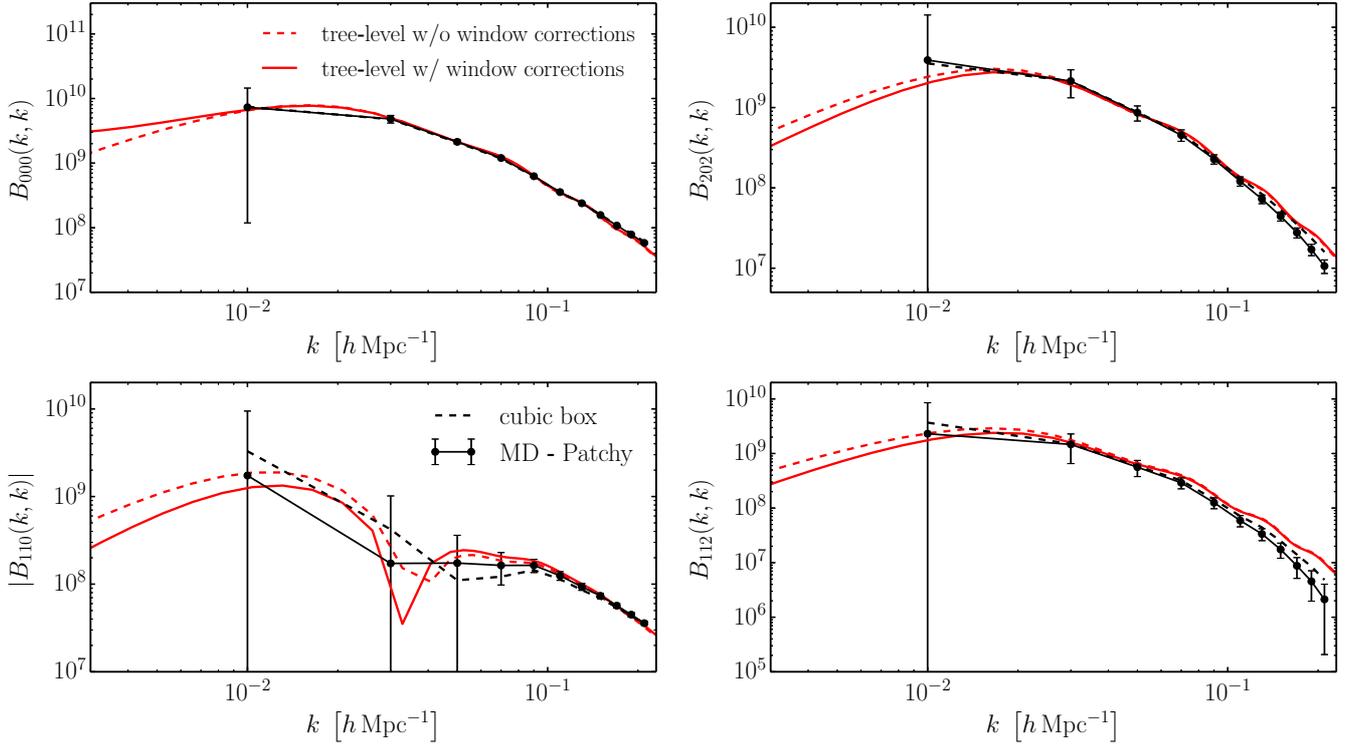}
	\caption{
    Diagonal bispectrum multipoles with (solid lines) and without (dashed lines) the survey geometry effect, for $B_{000}$ (upper left), $B_{110}$ (bottom left), $B_{202}$ (upper right), and $B_{112}$ (bottom right).
    In the context of simulations, we compare the measurements from the MD-Patchy mocks
    (black solid lines with error bars)
    with the mean values in the 200 cubic box simulations at $z=0.576$ that has the same galaxy distribution as the MD-Patchy mocks (black dashed lines).
    In the framework of perturbation theories, we take into account the survey geometry effect 
    using equation~(\ref{Eq:masked_bispectrum}),
    and plot the tree-level solutions with (red solid) and without (red dashed) the survey geometry effect.
     This figure shows that the survey geometry effect is noticeable mainly on scales larger than
    $k=0.01\hk$ for $B_{000}$, $k=0.1\hk$ for $B_{110}$, $k=0.02\hk$ for $B_{202}$, and $k=0.03\hk$ for $B_{112}$,
    and will be smaller on small scales, 
    even though Fig.~\ref{fig:bk_window_ratio} will show that the effective difference compared to the $1\sigma$ error can be still substantial even on small scales.
	}
	\label{fig:bk_window}
\end{figure*}

In the configuration-space analysis, it is not common to normalize the three-point function using the factor $I$ (equation~\ref{Eq:normalization}) but the window function monopole $Q_{000}(r_1,r_2)$. Equation~(\ref{Eq:zetalllMask}) then becomes
\begin{eqnarray}
	\frac{\left\langle \widehat{\zeta}_{\ell_1\ell_2L}(r_1,r_2)\right\rangle_{\rm model}}{Q_{000}(r_1,r_2)}
	\hspace{-0.25cm}&=&\hspace{-0.25cm} \zeta_{\ell_1\ell_2L}(r_1,r_2)\nonumber \\
	\hspace{-0.25cm}&+&\hspace{-0.25cm} \left[  \mbox{the other correction terms}  \right].
\end{eqnarray}
Thus, we can remove the window effect from the first term, but the other correction terms remain. As far as we measure $\zeta_{\ell_1\ell_2L}$ via the FFT-based estimator (equation~\ref{Eq:reduced_xi_estimator}) and fit it to a theoretical model, the above expression is necessary~\footnote{By solving the survey edge correction equation, we can correct for the survey geometry effect in configuration space only using the measured quantities themselves. This prescription has been developed by~\cite{Slepian:2015qza,Slepian:2017lpm}. If we focus only on the $L=0$ mode, equation~(\ref{Eq:zetaMask}) divided by $Q_{000}$ reduces to equation~(32) in~\cite{Slepian:2015qza}.}.

Finally, we obtain the masked bispectrum multipoles through the Hankel transform of $\langle \widehat{\zeta}_{\ell_1\ell_2L}\rangle_{\rm model}$:
\begin{eqnarray}
	&& \hspace{-0.4cm}\left\langle \widehat{B}_{\ell_1\ell_2L}(k_1,k_2) \right\rangle_{\rm model} \nonumber \\
	\hspace{-0.25cm}&=&\hspace{-0.25cm} 
	(-i)^{\ell_1+\ell_2}(4\pi)^2 \int dr_1 r_1^2 \int dr_2 r_2^2  \nonumber \\
	\hspace{-0.25cm}&\times&\hspace{-0.25cm}
	j_{\ell_1}(k_1r_1) j_{\ell_2}(k_2r_2) 
	\left\langle \widehat{\zeta}_{\ell_1\ell_2L}(r_1,r_2)\right\rangle_{\rm model}.
	\label{Eq:masked_bispectrum}
\end{eqnarray}
This masked bispectrum model is used to compare the theory with the measured bispectrum estimator during the likelihood fitting. There are two points to note about the above expression. The first one is that the masked bispectrum multipoles are proportional to $1/I$ (equation~\ref{Eq:normalization}) through the window function multipoles. Therefore, comparing the theory with the measurements, the value of $I$ no longer matters as mentioned in Section~\ref{Sec:Estimator}. The second one is that the masked bispectrum multipoles are derived using the local plane-parallel approximation in equation~(\ref{Eq:zeta_obs_theory}). This means that we can take into account the local plane-parallel approximation through the masked bispectrum multipoles as mentioned in Section~\ref{Sec:NewDecompositionFormalsim}, even though we can use the global plane-parallel formalism to compute the three-point functions that are necessary to calculate the masked three-point function in equation~(\ref{Eq:zetaMask}).

\subsection{Comparison with simulations}
\label{Sec:ComparisonWithSimulations}

\begin{figure}
	\includegraphics[width=1.0\columnwidth]{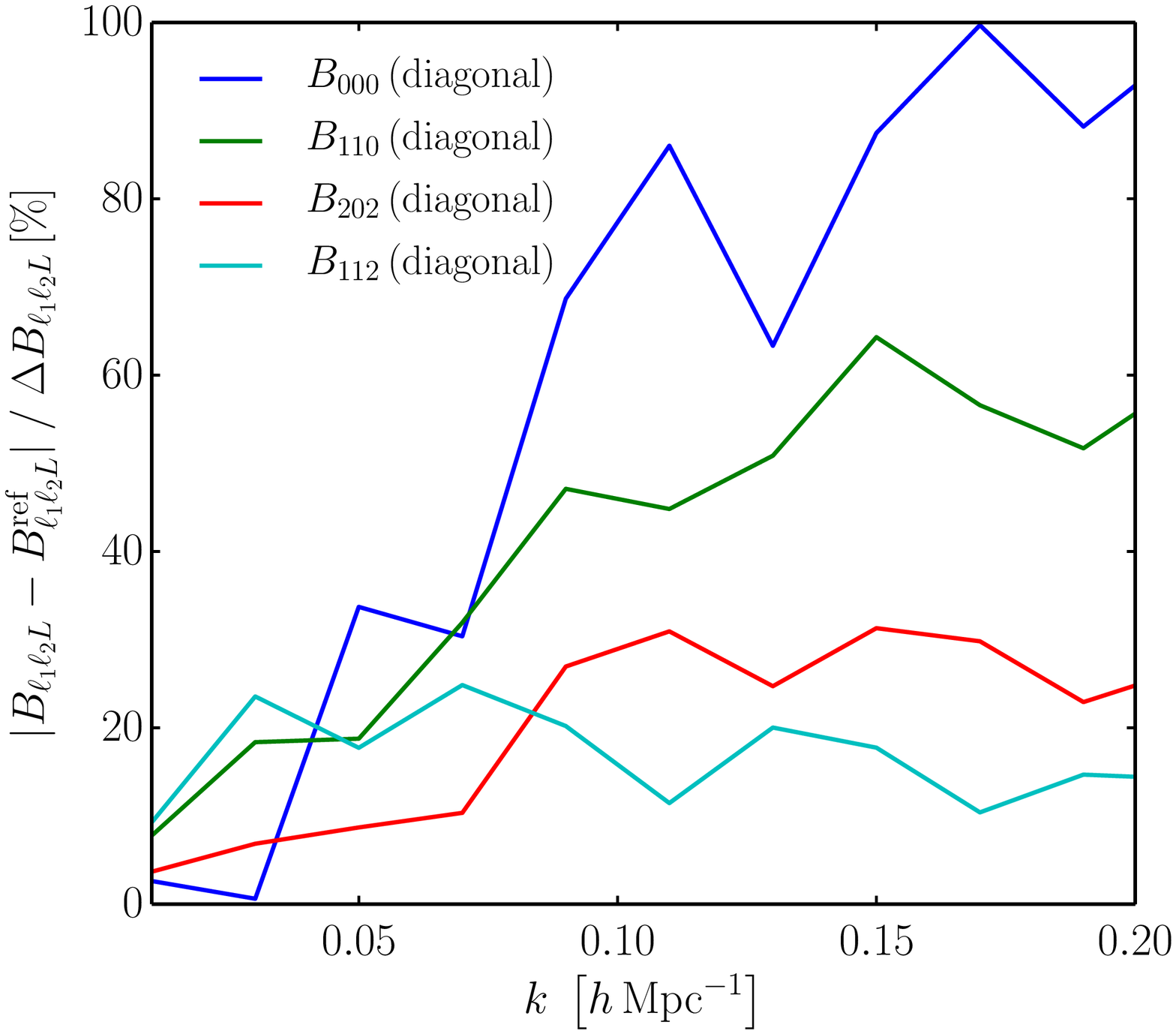}
	\caption{
    Ratios of the difference between the bispectrum multipoles with and without the survey window corrections, $B_{\ell_1\ell_2L} - B_{\ell_1\ell_2L}^{\rm ref}$ to the $1\sigma$ errors estimated from the MD-Patchy mocks, $\Delta B_{\ell_1\ell_2L}$.
    In this figure, we compute the bispectrum multipoles using the tree-level solution,
    and the reference bispectrum multipoles, $B_{\ell_1\ell_2L}^{\rm ref}$, are the true values  	 without including the survey window effect.
	This figure shows that although the absolute difference shown in Fig.~\ref{fig:bk_window} is small on small scales, the effective difference compared to the $1\sigma$ error reaches $50-100\%$ for $B_{000}$ and $B_{110}$, indicating that 
    the window corrections can not be ignored in clustering analysis even on small scales. 
	}
	\label{fig:bk_window_ratio}
\end{figure}

A direct way to estimate the survey geometry effect on the bispectrum measurements is to compare two types of simulations. The first one is the MD-Patchy mocks that reproduce the survey geometry of the BOSS galaxy sample. The second one is cubic box simulations which have the same galaxy description as the MD-Patchy mock catalogues with a volume of $( 2.5\hMpc )^3$. 
These box simulations provide the bispectrum measurement under the periodic boundary condition, without an artifact due to non-uniform survey footprint, and therefore, can be straightforwardly compared with the analytic bispectrum derived from perturbation theories.
The MD-Patchy mocks include the light-cone evolution of galaxy clustering that are produced with
different redshift snapshots of the box simulation.
As an approximation, we ignore the redshift dependence and compare one snapshot at $z=0.576$ with the MD-Patchy mocks.
To validate this approximation, we also measured the bispectra at three redshifts,
$z=0.466$, $z=0.505$, and $z=0.533$; we find that there is no significant difference among these snapshots, and hence, the evolution of galaxy clustering does not produce a large change of the shape of the bispectrum multiples in the redshift range $0.466< z < 0.576$. 
We expect from this result that the difference between the box simulations and the MD-Patchy mocks mainly arises from the survey geometry effect.
We note that we do not take account of the discreteness effect due to binning which may introduce systematic bias in the measured bispectrum at low-k at a few percent level. 
We defer a more detail analysis investigating this effect for future works.

In the box simulation, we apply the global-plane parallel approximation and choose the $z$-axis as the LOS direction. Then, we measure the bispectrum multipoles via the estimator given by equation~(\ref{Eq:simulation_estimator}).
	
Fig.~\ref{fig:bk_window} shows four diagonal bispectrum multipoles: $B_{000}$ (upper left), $B_{110}$ (bottom left), $B_{202}$ (upper right), and $B_{112}$ (bottom right). The black dashed lines denote the means of the measured bispectra in $200$ realizations of the box simulation at redshift $z=0.576$. The black solid lines with error bars in Fig.~\ref{fig:bk_window} are the results from the MD-Patchy mocks, which are the same as shown in Fig.~\ref{fig:bk}. 
We find clear differences between the box simulations and the MD-Patchy mocks 
at scales larger than $k=0.01\hk$ for $B_{000}$, $k=0.1\hk$ for $B_{110}$, $k=0.02\hk$ for $B_{202}$, and $k=0.03\hk$ for $B_{112}$.
For the $L=0$ mode ($B_{000}$ and $B_{110}$),
the MD-Patchy mocks including the survey window effect tend to be in agreement with the box mocks that provide the true measurements on small scales.
On the other hand, 
there is a small offset on small scales ($k\sim 0.2\hk$) in the $L=2$ mode ($B_{202}$ and $B_{112}$) between the two mocks, which could be due to the velocity calibrations made to the MD-Patchy mocks to match their quadrupole power spectrum to the measured quadrupole of BOSS DR12~\citep{Kitaura:2015uqa}.

The main purpose of this subsection is to show that our window function treatment with equation~(\ref{Eq:masked_bispectrum}) can indeed model the difference between the box simulations and the MD-Patchy mocks. To construct the bispectrum multipole models with the window function effect, we model the true bispectrum multipoles using second-order perturbation theory, the so-called tree-level solution (for details, see Appendix~\ref{Ap:WindowCorrections}). We compute the tree-level bispectra with the fiducial cosmological parameters mentioned in Section~\ref{Sec:Data} at $z=0.576$, where this parameter set corresponds to the box simulation shown in Fig.~\ref{fig:bk_window}. In this model, only two nuisance parameters are necessary to describe the clustering of a sample of galaxies: $b_1$ and $b_2$, the first and second order Eulerian bias parameters. We determine these bias parameters by performing a standard minimum $\chi^2$ analysis by comparing the four bispectrum multipoles, $B_{000}$, $B_{110}$, $B_{202}$, and $B_{112}$, predicted by the theory with those measured from the box simulation resulting in the best fitting values: $b_1 = 1.69$ and $b_2=1.89$. 
Notice that, since we adopt this model just for the purpose of reproducing the box simulation results at large scales, we do not interpret values of $b_1$ and $b_2$ as the physical galaxy bias parameters.
We compute the three-point function multipoles $\zeta_{\ell_1\ell_2L}(r_1,r_2)$ through the Hankel transform  given in equation~(\ref{Eq:HankelTransform}).

We adopt the simplest expression of the masked three-point function that only considers the first term in equation~(\ref{Eq:zetalllMask}):
\begin{eqnarray}
	\left\langle \widehat{\zeta}_{\ell_1\ell_2L}(r_1,r_2) \right\rangle_{\rm model} = Q_{000}(r_1,r_2) \zeta_{\ell_1\ell_2L}(r_1,r_2).
	\label{Eq:MaskZeta}
\end{eqnarray}
This approximation is adequate in the BOSS analysis because we have checked in Appendix~\ref{Ap:WindowCorrections} that the other additional corrections (the second line of equation~(\ref{Eq:zetalllMask})) produces a negligibly small change in the predictions of the bispectrum multipoles, $\lesssim 30\%$ of the $1\sigma$ errors estimated from the MD-Patchy mocks.

The red-solid and red-dashed lines in Fig.~\ref{fig:bk_window} show the tree-level solutions with and without the window corrections, respectively. 
The tree-level solution including our survey window corrections can model well the observed survey geometry effect.
In particular, the perturbation theory approach explains a characteristic feature that the survey window effect on $B_{110}$ appears even up to $k=0.1\hk$,
which is consistent with the result from the two mocks.

While the survey geometry effect on the bispectrum becomes smaller on small scales as shown in Fig.~\ref{fig:bk_window}, 
it is important to estimate the impact of the effect on clustering analysis more quantitatively.
To do so, we compute the ratios of the differences between the bispectrum multipoles with and without the window effect, $B_{\ell_1\ell_2L}- B_{\ell_1\ell_2L}^{\rm ref}$, to the $1\sigma$ error estimated from the MD-Patchy mocks, $\Delta B_{\ell_1\ell_2L}$, in Fig.~\ref{fig:bk_window_ratio}. 
Here, the reference bispectrum is the one without the window effect,
and we consider four diagonal bispectrum multipoles: $B_{000}$ (blue), $B_{110}$ (green), $B_{202}$ (red), and $B_{112}$ (cyan). 
The ratios shown in this figure increase at smaller scales,
because the $1\sigma$ errors decrease with increasing the number of modes,
i.e., with decreasing scales, faster than the window effect is reduced.
Remarkably, the ratios for $B_{000}$ and $B_{110}$ reach $\sim100\%$ and $\sim 60\%$
at $k=0.2\hk$, respectively. This fact indicates that we can not ignore the window effect 
in clustering analysis even on small scales.

Fig.~\ref{fig:bk_window} also clarifies the scales where the tree-level approximation breaks down. For $B_{000}$ and $B_{202}$, the tree-level solutions work well up to $k\sim0.1\hk$ at $z\sim0.5$ (compare the red dashed lines with the black dashed lines), while for $B_{110}$ and $B_{112}$, they cannot explain the simulation results at even larger scales. These results indicate that to model higher bispectrum multipoles, or to use small-scale information on the bispectrum, we need to go beyond the tree-level approximation. We leave improvements on the modeling of the bispectrum for future work (e.g.,~\citealt{Hashimoto:2017klo,Yamamoto:2016anp,Nan:2017oaq}).

\section{CONCLUSIONS}
\label{Sec:Conclusion}

This work aims to efficiently extract information of the RSD and AP effects from the three-point statistics to improve cosmological parameters constraints.

We have presented a new type of decomposition of the three-point statistics into the TripoSH basis (equation~\ref{Eq:TripoSH}) with zero total angular momentum, which makes use of the fact that density fluctuations in the Universe are expect to satisfy homogeneity, isotropy, and parity symmetry. In the relevant expansion coefficients for the bispectrum $B_{\ell_1\ell_2L}(k_1,k_2)$ (equation~\ref{Eq:reduced_ones}), the $L$ mode is an analog of the $\ell$ mode of the power spectrum multipoles, $P_{\ell}$, expanded in Legendre polynomials with respect to the LOS, and hence, the $L\neq0$ modes can never be generated unless in the presence of the RSD or AP effects. The corresponding coefficients of the three-point function $\zeta_{\ell_1\ell_2L}(r_1,r_2)$ can then be computed through a two-dimensional Hankel transform (equation~\ref{Eq:HankelTransform}).

We have presented the bispectrum and three-point function estimators within our new decomposition formalism 
both for the global plane-parallel approximation (i.e., for idealized periodic-box simulations, equation~\ref{Eq:simulation_estimator}) and for the local plane-parallel approximation (i.e., for observed galaxy data, equations~\ref{Eq:bi_estimator_FFTs} and~\ref{Eq:reduced_xi_estimator}). The resulting estimators can be computed using FFTs, which allows a complexity of $\mathcal{O}(N\log N)$, where N represents the number of grid cells used in the FFT.

Using our estimators,  we have measured the bispectrum multipoles from the CMASS NGC sample drawn from BOSS DR12 and the associated MD-Patchy mocks. We find an excellent agreement between the results from the mocks and the galaxy sample on large scales ($k\gtrsim0.1\hk$). On smaller scales ($k\lesssim0.1\hk$), they start to depart from each other beyond the $1\sigma$ error.

We have investigated the covariance matrices of the bispectrum multipoles (Fig.~\ref{fig:cov_bk}), calculated from the MD-Patchy mocks. Since the bispectrum multipoles are characterized by two scales, $k_1$ and $k_2$, their covariance matrices depend on four variables: $k_1$, $k_2$, $k'_1$, and $k'_2$. As far as we see the $k_1=k'_1$ (or $k_1=k'_2$) case, there are strong correlations between the bispectrum multipoles, even if $k_2\neq k'_2$ (or $k_2\neq k'_1$). Remarkably, there are weak correlations between the diagonal elements of the bispectrum multipoles estimated at two different scales,
i.e., between $B_{\ell_1\ell_2L}(k,k)$ and $B_{\ell_1\ell_2L}(k',k')$, indicating that they include nearly independent information at different scales. 

We have estimated the cumulative ${\rm S/Ns}$ of the amplitude of density fluctuations from the CMASS NGC sample (Fig.~\ref{fig:TotalSN}). In the analysis only considering the power spectrum monopole, there is the well-known nonlinear degradation of the ${\rm S/N}$ by $\sim80\%$ at $k_{\rm max}=0.1\hk$ and $\sim60\%$ at $k_{\rm max}=0.2\hk$ compared to the Gaussian prediction due to the leakage of the information to higher order statistics. The joint analysis of the power spectrum and bispectrum monopoles recovers the ${\rm S/N}$ up to $95\%$ at $k=0.1\hk$ and $75\%$ at $k=0.2\hk$. We find that the joint analysis using the diagonal bispectrum monopole well recovers the ${\rm S/N}$ up to $90\%$ at $k=0.1\hk$ and $70\%$ at $k=0.2\hk$ because of their nearly independent information at different scales.

We have also computed the ${\rm S/Ns}$ of the diagonal bispectrum multipoles themselves (Fig.~\ref{fig:SN}). For each of the $L=0$ and $2$ modes, the lowest order of the bispectrum multipoles, $B_{000}$ ($L=0$) or $B_{202}$ ($L=2$), yields the highest ${\rm S/N}$. 
The one of the most remarkable result of this paper is that, we, for the first time, detected the $L=2$ mode, which is an analog of the power spectrum quadrupole: we measure the lowest order $L=2$ mode, $B_{202}$, with a significance of $14\sigma$ and also the next leading order, $B_{112}$, with a significance of $6\sigma$.

The finding that most of the bispectrum information is contained in the diagonal terms could significantly simplify the bispectrum analysis. The measurements of the diagonal elements of a few bispectrum multipoles in both $L=0$ and $L=2$ modes would allow extracting most of the higher order information contained in the bispectrum.

We present a formalism for how to include the survey window function in the bispectrum model, before comparing it with the measured bispectrum. This represents the first self-consistent treatment of the survey window function in a bispectrum analysis. The survey geometry effect appears mainly on large scales and gets close to zero at small scales (Fig.~\ref{fig:bk_window}) as expected. Nevertheless, the survey geometry effect can significantly affect the bispectrum measurements even at small-scales, because these survey geometry effects are comparable to the statistical uncertainties, which is also reduced towards small-scales (Fig.~\ref{fig:bk_window_ratio}). This fact emphasizes the need to include the survey geometry effect in any statistical analysis.

This paper has demonstrated that the TripoSH formalism can be used to measure the bispectrum in large-scale structure datasets, which complements the now well-established power spectrum analysis. The formalism can be trivially extended to higher-order statistics, like the four-point function and the trispectrum, by use of poly-polar spherical harmonic (PolypoSH) decomposition. We hope that the formalism presented in this paper will become a standard method for analyzing higher-order statistics.

\section*{Acknowledgements}

NSS acknowledges financial support from Grant-in-Aid for JSPS Fellows (Nos. 28-1890). Numerical computations were carried out on Cray XC50 at Center for Computational Astrophysics, National Astronomical Observatory of Japan. This work was supported in part by JSPS KAKENHI Grant Number JP15H05896.
FB is a Royal Society University Research Fellow. H-JS is supported by the U.S.~Department of Energy, Office of Science, Office of High Energy Physics under Award Number DE-SC0014329. NSS would like to thank Eiichiro Komatsu and Max Planck Institute for Astrophysics for their warm hospitality during the middle stage of this work.
NSS and SS also thank the Yukawa Institute for Theoretical Physics at Kyoto University. Discussions during the YITP workshop YITP-T-17-03 on ``CosKASI-ICG-NAOC-YITP'' were used to complete this work.

\bibliographystyle{mnras}
\bibliography{ms} 

\appendix

\section{Useful identities}
\label{Ap:UsefulIdentities}
Here we summarize the identities used for the derivations in this paper.

Throughout this paper, we use a normalized spherical harmonics, $y_{\ell}^m$, defined as
\begin{eqnarray}
	y_{\ell}^m(\hat{x}) \equiv \sqrt{\frac{4\pi}{2\ell+1}}\, Y_{\ell}^m(\hat{x}),
\end{eqnarray}
where its leading order becomes unity: $y_{0}^0 = 1$.

The relation between Legendre polynomials and spherical harmonics is given by
\begin{eqnarray}
	{\cal L}_{\ell}(\hat{x}\cdot\hat{y}) = \sum_{m}\, y_{\ell}^{m}(\hat{x})\,y_{\ell}^{m*}(\hat{y}).
	\label{Eq:LYlm}
\end{eqnarray}

The relation between associated Legendre polynomials and spherical harmonics is 
\begin{eqnarray}
	y_{\ell}^m(\theta,\varphi) = (-1)^{(m-|m|)/2}\sqrt{\frac{(\ell-|m|)!}{(\ell+|m|)!}} 
	{\cal L}_{\ell}^{|m|}(\cos \theta) \mathrm{e}^{\mathrm{i}m\varphi}.
	\label{Eq:YL}
\end{eqnarray}

The addition rule of spherical harmonics:
\begin{eqnarray}
	\hspace{-0.53cm}&&y_{\ell_1}^{m_1}(\hat{x})\, y_{\ell_2}^{m_2}(\hat{x}) \nonumber \\
	\hspace{-0.23cm}&=&\hspace{-0.23cm}
	\sum_{\ell_3m_3} (2\ell_3+1) \left( \begin{smallmatrix} \ell_1 & \ell_2 & \ell_3 \\ 0 & 0 & 0 \end{smallmatrix}  \right)
	\left( \begin{smallmatrix} \ell_1 & \ell_2 & \ell_3 \\ m_1 & m_2 & m_3 \end{smallmatrix}  \right) y_{\ell_3}^{m_3*}(\hat{x}). 
	\label{Eq:YY2Y}
\end{eqnarray}

Five Wigner $3$-$j$ symbols can generate the Wigner $9$-$j$ symbol as follows:
\begin{eqnarray}
	&&\sum_{m_4m_5m_6m_7m_8m_9}(-1)^{\sum_{i=4}^9 (\ell_i - m_i)} \nonumber \\
	&\times&
	\left( \begin{smallmatrix} \ell_4 & \ell_1 & \ell_7 \\ m_4 & -m_1 & m_7 \end{smallmatrix}  \right)
	\left( \begin{smallmatrix} \ell_5 & \ell_4 & \ell_6 \\ -m_5 & -m_4 & -m_6 \end{smallmatrix}  \right) \nonumber \\
	&\times&
	\left( \begin{smallmatrix} \ell_6 & \ell_3 & \ell_9 \\ m_6 & -m_3 & m_9 \end{smallmatrix}  \right)
	\left( \begin{smallmatrix} \ell_7 & \ell_9 & \ell_8 \\ -m_7 & -m_9 & -m_8 \end{smallmatrix}  \right)
	\left( \begin{smallmatrix} \ell_8 & \ell_2 & \ell_5 \\ m_8 & -m_2 & m_5 \end{smallmatrix}  \right) \nonumber \\
	&=&(-1)^{\ell_2+\ell_5+\ell_8} 
	\left( \begin{smallmatrix} \ell_1 & \ell_2 & \ell_3 \\ m_1 & m_2 & m_3 \end{smallmatrix}  \right) 
	\left\{ \begin{smallmatrix} \ell_1 & \ell_2 & \ell_3 \\   \ell_4 & \ell_5 & \ell_6 \\   \ell_7 & \ell_8 & \ell_9 \end{smallmatrix}  \right\}.
	\label{Eq:Wigner9j}
\end{eqnarray}

\section{Choice of line-of-sight directions}
\label{Ap:LOS}

In the main text, when measuring the bispectrum multipoles from observed galaxies, we apply the local plane-parallel approximation, $\hat{x}_1\approx\hat{x}_2\approx\hat{x}_3$, and choose $\hat{x}_3$ as the LOS direction. However, as the approximation is not exactly satisfied in the bispectrum measurement, the $L\geq 2$ modes of the measured bispectrum multipoles may depend on the choice of the LOS direction: $\hat{x}_1$, $\hat{x}_2$, or $\hat{x}_3$, while the $L=0$ mode does not.

\begin{figure}
	\includegraphics[width=\columnwidth]{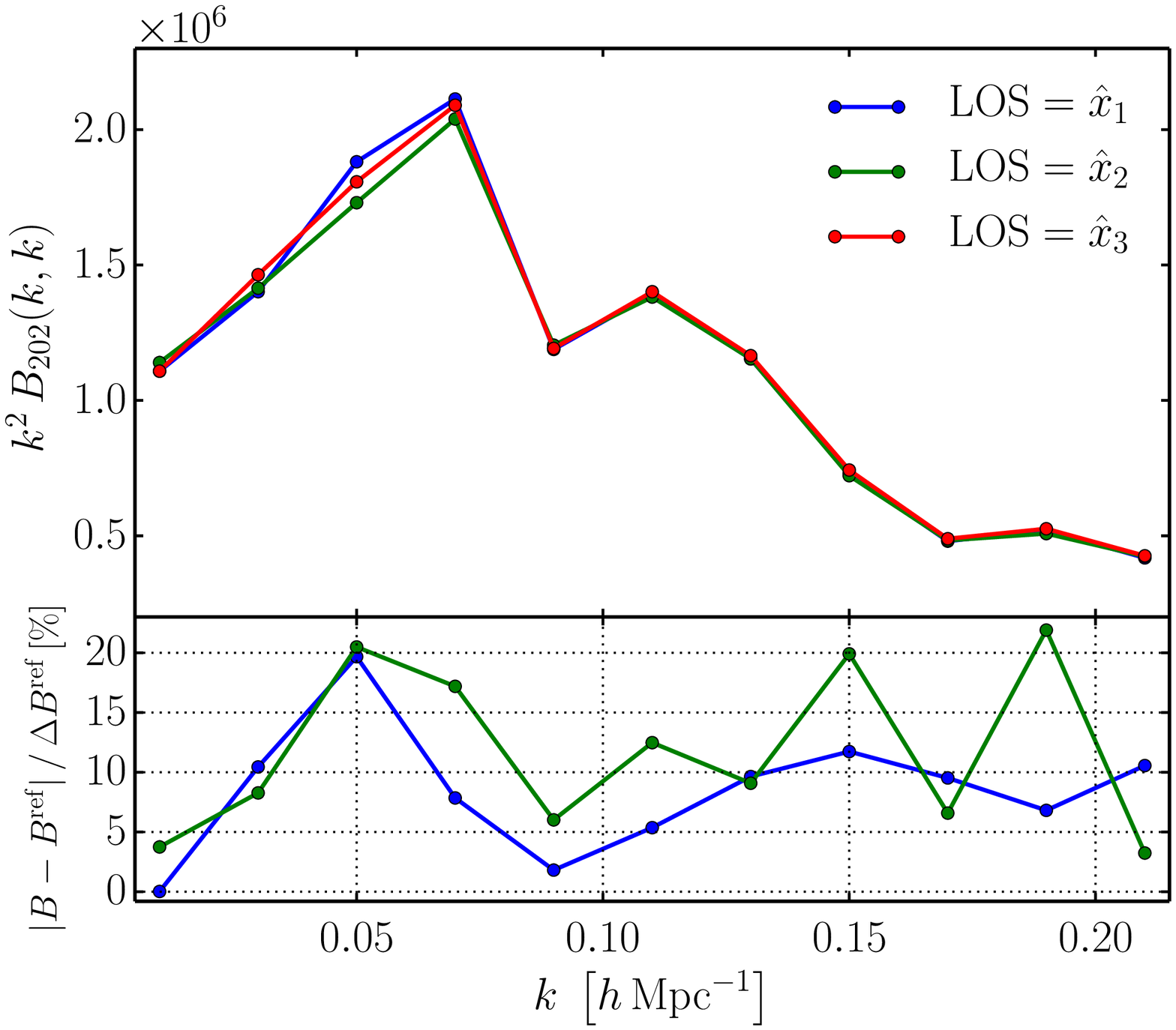}
	\caption{
		A comparison of the bispectrum measurements using three different LOS directions from CMASS NGC.
		In the top part, we measure the lowest order of the $L=2$ mode of the bispectrum multipoles, $B_{202}(k,k)$, 
		by choosing $\hat{x}_1$, $\hat{x}_2$, and $\hat{x}_3$ as the LOS direction.
		In the bottom part, we show the ratios of the differences between the measured bispectra and a reference bispectrum to the $1\sigma$ error, $\Delta B^{\rm ref}$, on the reference one.	
		This figure shows that the dependence of the choice of the LOS directions on the bispectrum measurement is negligibly small in the BOSS analysis, 
		$\lesssim20\%$ of $\Delta B^{\rm ref}$.
	}
	\label{fig:LOS}
\end{figure}

This point is illustrated in Fig.~\ref{fig:LOS} which shows the lowest order of the $L=2$ mode, $B_{202}(k,k)$,  measured using three different LOS directions, $\hat{x}_1$ (blue), $\hat{x}_2$ (green), and $\hat{x}_3$ (red), from the CMASS NGC sample. While the top part plots the bispectrum measurements, the bottom shows the ratios of the differences between the measured bispectra and a reference bispectrum $B^{\rm ref}$ to the $1\sigma$ error $\Delta B^{\rm ref}$ on the reference one. Here, the reference bispectrum is computed using $\hat{x}_3$ as the LOS direction (Section~\ref{Sec:Estimator}), and its $1\sigma$ error is estimated from the $600$ MD-Patchy mocks (Section~\ref{Sec:CovarianceMatrix}). We find from this figure that the dependence of the choice of the LOS directions on the bispectrum measurement is less than about $20\%$ of the $1\sigma$ error, and hence that it is negligible in the BOSS analysis. We expect that the difference among the LOS directions, to some extent, can be explained by the survey window corrections (Section~\ref{Sec:SurveyWindowFunctions}), because the survey window function should also depend on the choice of the LOS directions.

\section{Mass assignment schemes}
\label{Ap:MassAssignmentFunction}

Measuring a density field from a galaxy sample requires the interpolation of functions on a regular grid in position space. In this section, we provide a comparison of several interpolation schemes, the nearest grid point (NGP), cloud-in-cell (CIC), and triangular-shaped cloud (TSC) assignment schemes~\citep{HockneyEastwood1981}, with two different number of grid-cells, $N_{\rm grid}=512^3$ and $N_{\rm grid}=1024^3$. We focus especially on the diagonal bispectrum monopole, $B_{000}(k,k)$, and measure it from the CMASS NGC galaxy sample by placing the CMASS NGC galaxies in a cubic box with a volume of $(3.5\hGpc)^3$. The Nyquist frequency then becomes $k_{\rm N}=0.46\hk$ for $N_{\rm grid}=512^3$ and $k_{\rm N}=0.92\hk$ for $N_{\rm grid}=1024^3$. 

\begin{figure}
	\includegraphics[width=\columnwidth]{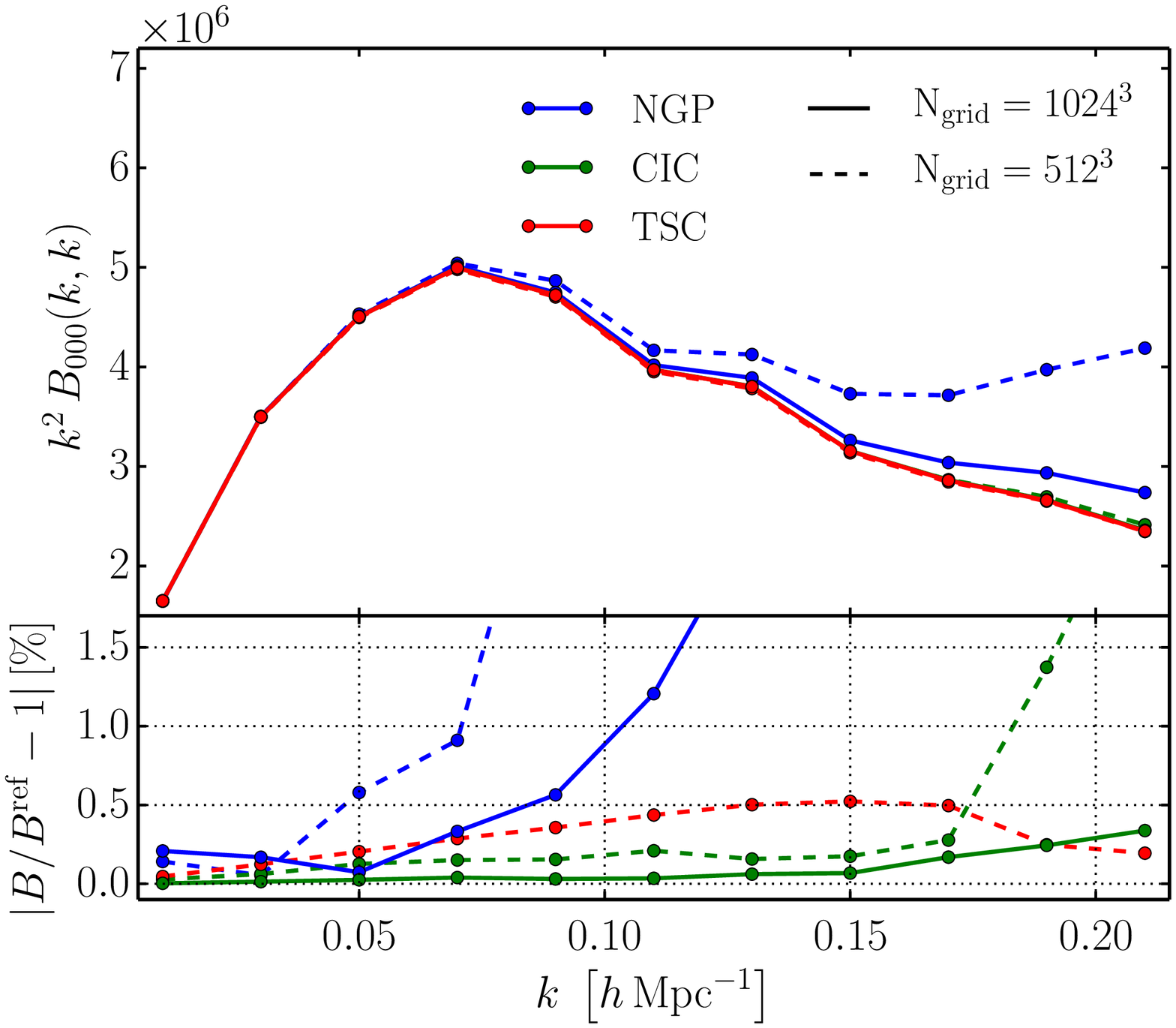}
	\caption{
		A comparison of the effect of the mass assignment function on the measurement of the diagonal bispectrum monopole, $B_{000}(k,k)$.
		In the top part, we measure $B_{000}(k,k)$ using the NGP, CIC, and TSC schemes with two different numbers of grid-cells, $N_{\rm grid}=512^3$ and $N_{\rm grid}=1024^3$, 
		from the CMASS NGP sample.
		The CMASS NGC galaxies are placed in a cubic box with a volume of $(3.5\hGpc)^3$, and the Nyquist frequency
		is then $k_{\rm N}=0.46\hk$ for $N_{\rm grid}=512^3$ and $k_{\rm N}=0.92\hk$ for $N_{\rm grid}=1024^3$. 
		In the bottom part, we show the ratios of the measured bispectrum monopoles to a reference one $B^{\rm ref}$, computed using 
		the TSC scheme with a $1024^3$ grids.
		This figure shows that the TSC scheme achieves sub-percent accuracy, $\lesssim0.5\%$, up to the scale of interest, $k\lesssim0.21\hk$.
	}
	\label{fig:MassAssignment}
\end{figure}

Fig.~\ref{fig:MassAssignment} illustrates the difference among the NGP (blue), CIC (green), and TSC (red) schemes with $N_{\rm grid}=512^3$ (dashed) and $N_{\rm grid}=1024^3$ (solid). The top part plots the measurements using these schemes, and the bottom shows the ratios of the measured bispectrum monopoles to a reference bispectrum monopole, where the reference one is computed using the TSC assignment function with a $1024^3$ grid. The choice of assignment functions significantly affects the bispectrum measurement, compared to the power spectrum case~\citep[e.g., see Figure 2 in][]{Hand:2017pqn}. In particular, the NGP scheme has a bad convergence property when increasing $N_{\rm grid}$. As expected, higher-order interpolation schemes perform better. For the TSC scheme, doubling the number of grids per side, from $512$ to $1024$, produces a small change in the bispectrum measurement, within $0.5\%$, in a $k$-range of $0.01\hk<k<0.21\hk$.

\section{FKP weighting}
\label{Ap:FKPWeight}

While the FKP weight function $w_{\rm FKP}(\VEC{x})=1/\left[ 1 + \bar{n}(\VEC{x})P_0\right]$ was originally obtained by minimizing the fractional variance in the power spectrum under the assumption that density fluctuations are Gaussian~\citet{Feldman:1993ky, Scoccimarro:2000sn} showed that assuming Gaussian errors, the same weight function minimizes the variance of the higher-order correlation functions as well. To verify this in our decomposition formalism (Section~\ref{Sec:NewDecompositionFormalsim}), we compare two measurements of the bispectrum multipoles with and without the FKP weighting, with $P_0 = 10^4\,h^{-3}\,{\rm Mpc}^3$. We use the $600$ MD-Patchy mocks for CMASS NGC to estimate the mean value of the bispectrum measurements and their $1\sigma$ errors.

\begin{figure}
	\includegraphics[width=\columnwidth]{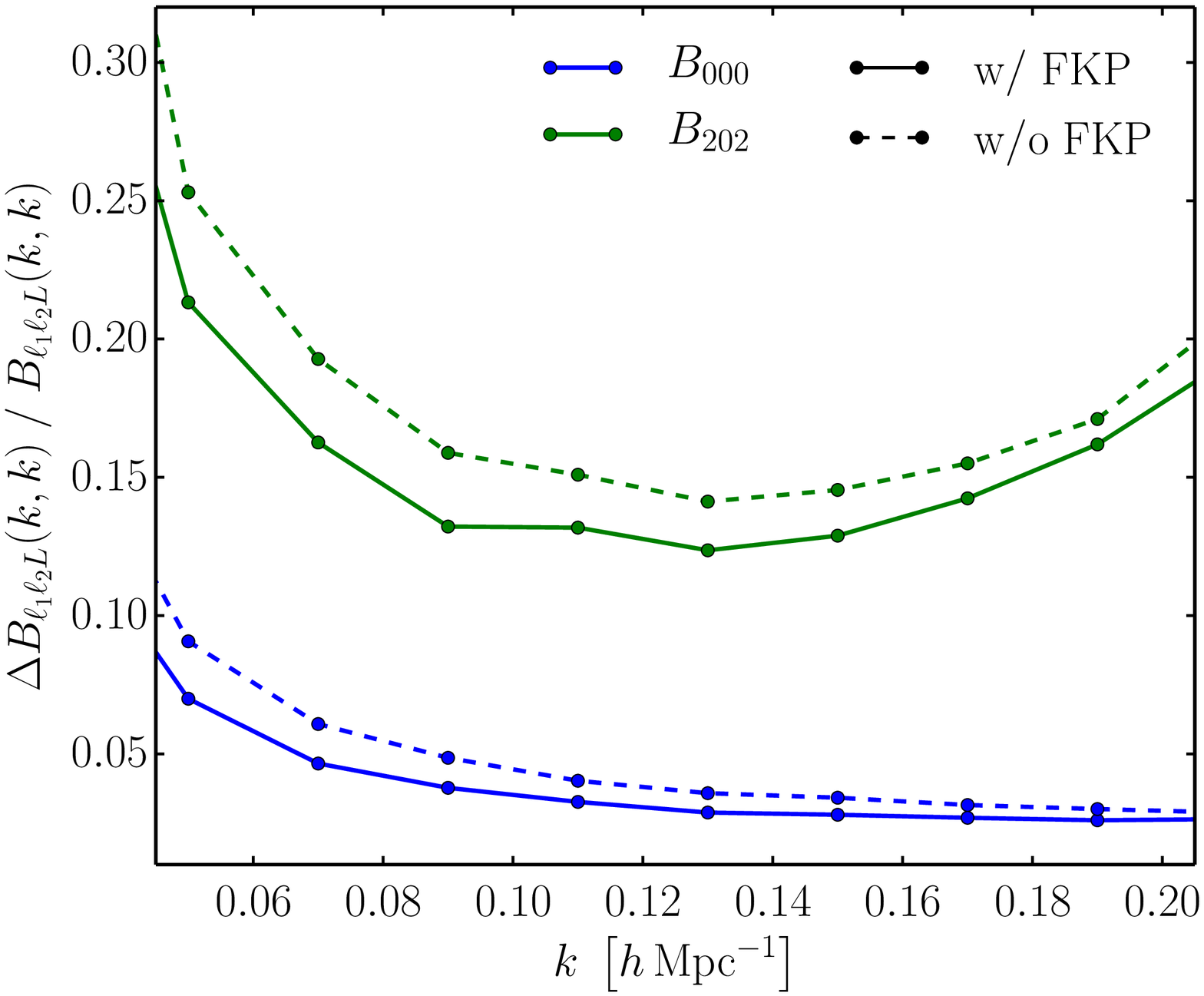}
	\caption{
		A comparison of the bispectrum measurements with (solid) and without (dashed) the FKP weighting from the CMASS NGC mocks,
		where both the lowest orders of the $L=0$ and $L=2$ modes, $B_{000}$ (blue) and $B_{202}$ (green), are computed.
		We plot the ratios of the $1\sigma$ error on the bispectrum measurements to the mean values of them.
		This figure shows that the FKP weighting decreases the errors by $\sim 15\%$ in the $k$-range $0.01\hk<k<0.2\hk$ in the BOSS analysis.
	}
	\label{fig:FKP}
\end{figure}

Fig.~\ref{fig:FKP} provides a comparison of the bispectrum measurements with (solid) and without (dashed) the FKP weighting, where we measure the lowest orders of the $L=0$ and $L=2$ modes, respectively: $B_{000}$ (blue) and $B_{202}$ (green), and we plots the ratios of the $1\sigma$ error on the bispectrum measurements to the mean values of them. We find from this figure that the FKP weighting, as expected, decreases the errors by $\sim 15\%$ in the $k$-range $0.01\hk<k<0.2\hk$ in the BOSS analysis.

\section{The Hartlap factor and the Percival factor}
\label{Ap:HartlapAndPercival}

In this appendix, we describe the Hartlap factor and the Percival factor.

Measuring the standard $\chi^2$ statistics in Section~\ref{Sec:SignalNoiseRatio}, it is necessary to compute the inverse of the covariance matrix.
As the estimated covariance matrix $\textbf{\textsf{C}}$ in equation~(\ref{Eq:Covariance}) is inferred from a set of mocks,
its inverse $\textbf{\textsf{C}}^{-1}$ is biased due to the limited number of realizations.
We account for this effect by rescaling the inverse covariance matrix as follows~\citep{Hartlap:2006kj}
\begin{eqnarray}
	\textbf{\textsf{C}}_{\rm Hartlap}^{-1} = \frac{N_{\rm mock}-N_{\rm bin}-2}{N_{\rm mock}-1} \textbf{\textsf{C}}^{-1},
\end{eqnarray}
where $N_{\rm bin}$ is the number of bins, and $(N_{\rm mock}-N_{\rm bin}-2)/(N_{\rm mock}-1)$ is the so-called Hartlap factor.
The standard $\chi^2$ value is then given by
\begin{eqnarray}
	\chi^2 = \left( \VEC{X}^{\rm obs} - \VEC{X}^{\rm theory} \right)^{\rm T}
	\textsf{\textbf{C}}^{-1}_{\rm Hartlap}
	\left( \VEC{X}^{\rm obs} - \VEC{X}^{\rm theory} \right).
\end{eqnarray}
where $\VEC{X}^{\rm obs}$ is the vector of the observed quantities, and $\VEC{X}^{\rm theory}$ is the vector of the corresponding theoretical predictions.

In addition to the Hartlap factor, we propagate the error in the covariance matrix to the error on parameters by scaling their variances by 
~\cite[equation~18 in][]{Percival:2013sga}:
\begin{eqnarray}
	M = \frac{1+B(N_{\rm bin}-N_{\rm p})}{1+A+B(N_{\rm p} +1)}
\end{eqnarray}
with
\begin{eqnarray}
	A \hspace{-0.25cm}&=&\hspace{-0.25cm} \frac{2}{\sqrt{\left( N_{\rm mock} - N_{\rm bin} - 1 \right)\left( N_{\rm mock} - N_{\rm bin} - 4 \right)}} \nonumber \\
	B \hspace{-0.25cm}&=&\hspace{-0.25cm} \frac{N_{\rm mock}-N_{\rm bin}-2}{\sqrt{\left( N_{\rm mock} - N_{\rm bin} - 1 \right)\left( N_{\rm mock} - N_{\rm bin} - 4 \right)}},
\end{eqnarray}
where $N_{\rm p}$ is the number of parameters.

\section{Window corrections}
\label{Ap:WindowCorrections}

We here provide a complementary study of the survey geometry effect on the bispectrum multipoles in the framework of perturbation theory.

In the second-order perturbation theory, the first and second-order kernels are~\citep{Scoccimarro:1999ed}
\begin{eqnarray}
	Z_1(\VEC{k}_1)  \hspace{-0.25cm}&=& \hspace{-0.25cm}  \left( b_1 + f\mu_1^2 \right) \nonumber\\
	Z_2(\VEC{k}_1,\VEC{k}_2)  \hspace{-0.25cm}&=& \hspace{-0.25cm} b_1 F_1(\VEC{k}_1,\VEC{k}_2) + \frac{b_2}{2} + f \mu^2 G_2(\VEC{k}_2,\VEC{k}_2) \nonumber \\
	 \hspace{-0.25cm}&+& \hspace{-0.25cm} \frac{f\mu k}{2} \left[ \frac{\mu_1}{k_1}Z_1(\VEC{k}_2) +  \frac{\mu_2}{k_2}Z_1(\VEC{k}_1)  \right]
\end{eqnarray}
where $\mu = \hat{k}\cdot\hat{n}$ with $\VEC{k}=\VEC{k}_1+\VEC{k}_2$, $\mu_i=\hat{k}_i\cdot\hat{n}$ for $i=1,2$, and
\begin{eqnarray}
	F_2(\VEC{k}_1,\VEC{k}_2) \hspace{-0.25cm}&=&\hspace{-0.25cm} 
	\frac{5}{7} + \frac{\mu_{12}}{2}\left( \frac{k_1}{k_2}+\frac{k_2}{k_1} \right) + \frac{2}{7} \mu_{12}^2 \nonumber \\
	G_2(\VEC{k}_1,\VEC{k}_2) \hspace{-0.25cm}&=&\hspace{-0.25cm} 
	\frac{3}{7} + \frac{\mu_{12}}{2}\left( \frac{k_1}{k_2}+\frac{k_2}{k_1} \right) + \frac{4}{7} \mu_{12}^2
\end{eqnarray}
with $\mu_{12}=\hat{k}_1\cdot\hat{k}_2$. The bispectrum is then given by
\begin{eqnarray}
	B(\VEC{k}_1,\VEC{k}_2,\VEC{k}_3,\hat{n}) \hspace{-0.25cm}&=&\hspace{-0.25cm}  2 Z_2(\VEC{k}_1,\VEC{k}_2) Z_1(\VEC{k}_1) Z_1(\VEC{k}_2) P_{\rm lin}(k_1)P_{\rm lin}(k_2) \nonumber \\
	                                 \hspace{-0.25cm}&+&\hspace{-0.25cm} {\rm cyc.},
\end{eqnarray}
where $P_{\rm lin}$ is the linear matter power spectrum, which is generated with \textsc{CLASS}~\citep{Lesgourgues:2011re} in this paper.

Computing the masked three-point function multipoles (equation~\ref{Eq:zetaMask}), in this appendix, we decide to use four three-point function multipoles and four window function multipoles: $\zeta_{000}$, $\zeta_{110}$, $\zeta_{202}$, and $\zeta_{112}$; $Q_{000}$, $Q_{110}$, $Q_{202}$, and $Q_{112}$. We ignore the terms expressed by a product of the $L=2$ modes of the three-point function and the window function, e.g., $\zeta_{202} Q_{202}$, because we found them to be negligible. Furthermore, we approximate the integral constraint (equation~\ref{Eq:IC}) as follows:
\begin{eqnarray}
	\bar{\zeta}\approx
	\frac{\int dr_1 r_1^2\int dr_2 r_2^2  Q_{000}(r_1,r_2)\zeta_{000}(r_1,r_2)}{\int dr_1 r_1^2\int dr_2 r_2^2\, Q_{000}(r_1,r_2)}.
\end{eqnarray}
The masked three-point function multipoles are then given by
\begin{eqnarray}
	\left\langle \widehat{\zeta}_{000}\right\rangle_{\rm model}
	\hspace{-0.25cm}&=&\hspace{-0.25cm} Q_{000}\left[\zeta_{000}-\bar{\zeta}\, \right] + \frac{1}{3}Q_{110}\zeta_{110} \nonumber \\
	\left\langle \widehat{\zeta}_{110}\right\rangle_{\rm model}
	\hspace{-0.25cm}&=&\hspace{-0.25cm} Q_{000}\zeta_{110} + Q_{110}\left[\zeta_{000}-\bar{\zeta}\, \right]  \nonumber \\
	\left\langle \widehat{\zeta}_{202}\right\rangle_{\rm model}
	\hspace{-0.25cm}&=&\hspace{-0.25cm}
	Q_{000}\zeta_{202} + \frac{1}{3}Q_{110}\zeta_{112} \nonumber\\
	\hspace{-0.25cm}&+&\hspace{-0.25cm}
	\frac{1}{3}Q_{112}\zeta_{110} + Q_{202}\left[ \zeta_{000}-\bar{\zeta}\, \right] \nonumber \\
	\left\langle \widehat{\zeta}_{112}\right\rangle_{\rm model}
	\hspace{-0.25cm}&=&\hspace{-0.25cm}
	Q_{000}\zeta_{112} + \frac{2}{5}Q_{110}\left[  \zeta_{022}+\zeta_{202}  \right] \nonumber \\
	\hspace{-0.25cm}&+&\hspace{-0.25cm}
	Q_{112}\left[ \zeta_{000}-\bar{\zeta}\, \right]+\frac{2}{5}\left[Q_{022}+Q_{202} \right]\zeta_{110},
	\label{Eq:MaskZeta2}
\end{eqnarray}
where for brevity, we abbreviated the dependence of two comoving distances, $r_1$ and $r_2$, on $\zeta_{\ell_1\ell_2L}(r_1,r_2)$ and $Q_{\ell_1\ell_2L}(r_1,r_2)$. By substituting these expressions into equation~(\ref{Eq:HankelTransform}), we can obtain the masked bispectrum multipoles. The purpose of this appendix is to compare the results from the above masked three-point function with the simplest approximation, $\langle \widehat{\zeta}_{\ell_1\ell_2L}\rangle_{\rm model} = Q_{000} \zeta_{\ell_1\ell_2L}$ (equation~\ref{Eq:MaskZeta}), used in the main text.

\begin{figure}
	\includegraphics[width=\columnwidth]{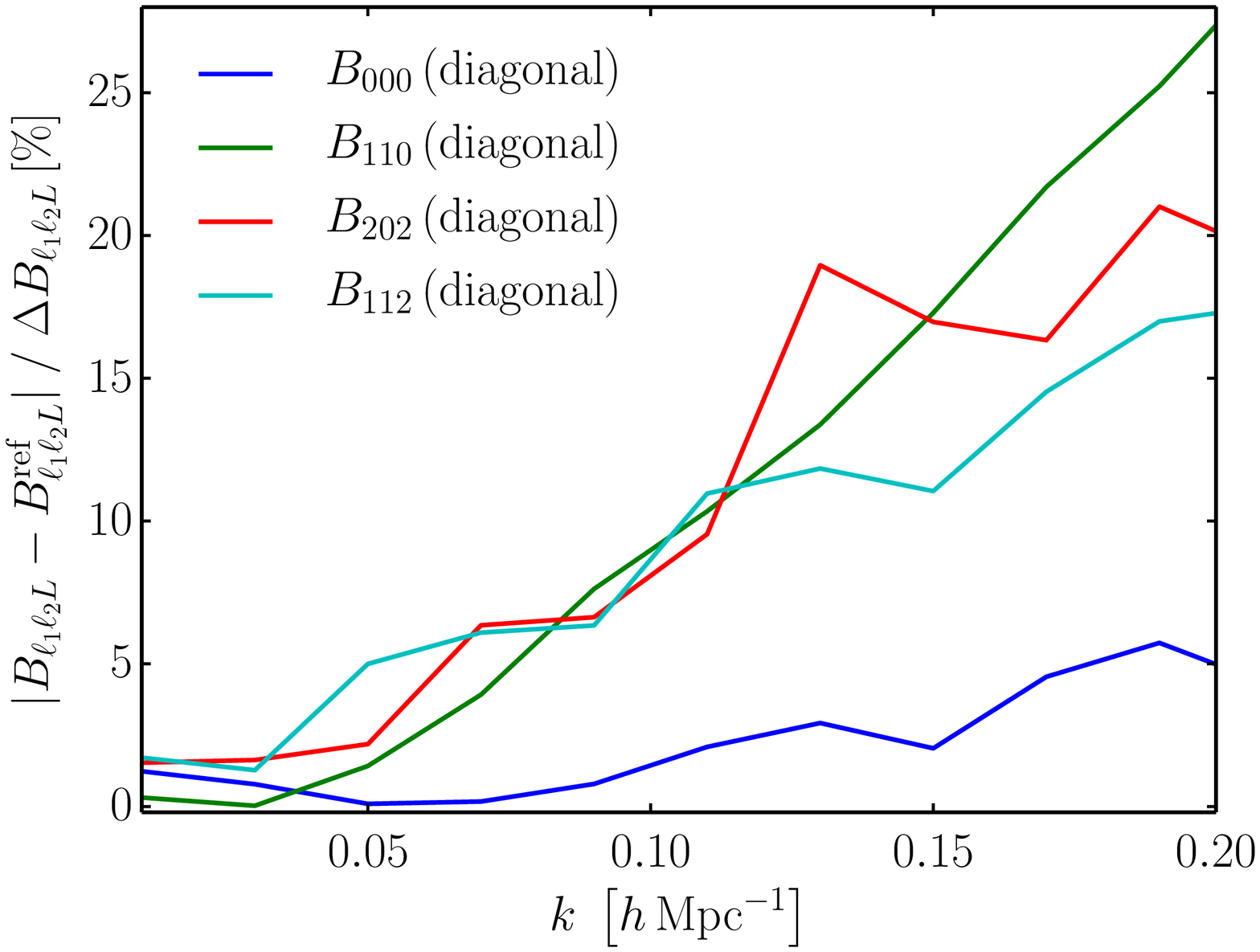}
	\caption{Ratios of the difference between the masked bispectrum multipoles computed using equations~(\ref{Eq:MaskZeta}) and (\ref{Eq:MaskZeta2})
	to the $1\sigma$ errors estimated from the MD-Patchy mocks,
	where the reference bispectrum multipoles, $B_{\ell_1\ell_2L}^{\rm ref}$, are computed from equation~(\ref{Eq:MaskZeta}).
	This figure shows that the difference is less than $\sim 30\%$ of the $1\sigma$ error,
	and hence, additional correction terms given by equation~(\ref{Eq:MaskZeta2}) to the window function effect are negligible in the BOSS analysis.
	}
	\label{fig:bk_WC}
\end{figure}

Fig.~\ref{fig:bk_WC} shows the ratios of the difference between equations~(\ref{Eq:MaskZeta}) and (\ref{Eq:MaskZeta2})
, $|B_{\ell_1\ell_2L} - B^{\rm ref}_{\ell_1\ell_2L}|$, to the $1\sigma$ errors estimated from the MD-Patchy mocks,
$\Delta B_{\ell_1\ell_2L}$, where the reference bispectrum multipoles are computed from equation~(\ref{Eq:MaskZeta}).
Both of the numerator $|B_{\ell_1\ell_2L} - B^{\rm ref}_{\ell_1\ell_2L}|$ 
and the denominator $\Delta B_{\ell_1\ell_2L}$ decrease toward smaller scales,
but $\Delta B_{\ell_1\ell_2L}$ reduces faster than $|B_{\ell_1\ell_2L} - B^{\rm ref}_{\ell_1\ell_2L}|$.
Consequently, the ratio between them becomes an increasing function.
We find that the difference is less than $\sim 30\%$ of the $1\sigma$ error, indicating that
the simplest approximation (equations~\ref{Eq:MaskZeta}) used in the main text is adequate for the BOSS analysis.

\bsp	
\label{lastpage}
\end{document}